\documentclass[review]{elsarticle}

\usepackage{lineno,hyperref}
\usepackage{moreverb}
\usepackage{bm,color,transparent,amsmath, amsmath,amssymb,amsthm,fancyhdr,mathrsfs,graphicx,setspace,stmaryrd,psfrag}
\usepackage[ruled,slide]{algorithm2e}
\newcommand\BibTeX{{\rmfamily B\kern-.05em \textsc{i\kern-.025em b}\kern-.08em
T\kern-.1667em\lower.7ex\hbox{E}\kern-.125emX}}

\DeclareMathOperator*{\A}{ \raisebox{-2pt}{$\mathlarger{\mathlarger{\mathlarger{\mathlarger{\mathlarger{\mathsf{A}}}}}}$} }

\journal{CMAME}

\begin{document}

\begin{frontmatter}

\title{Energy consistent framework for continuously evolving 3D crack propagation}

\author[gla]{\L{ukasz} Kaczmarczyk}
\ead{lukasz.kaczmarczyk@glasgow.ac.uk}
\author[gla]{Zahur Ullah\corref{cor}}
\ead{zahur.ullah@glasgow.ac.uk}
\author[gla]{Chris J. Pearce}
\ead{chris.pearce@glasgow.ac.uk}
\cortext[cor]{Corresponding author}
\address[gla]{School of Engineering, University of Glasgow, Glasgow, G12 8QQ, UK.}


\begin{abstract}
This paper presents a formulation for brittle fracture in 3D elastic solids within the context of configurational mechanics. The local form of the first law of thermodynamics provides a condition for equilibrium of the crack front. The direction of the crack propagation is shown to be given by the direction of the configurational forces on the crack front that maximise the local dissipation. The evolving crack front is continuously resolved by the finite element mesh, without the need for face splitting or the use of enrichment techniques. 
A monolithic solution strategy is adopted, solving simultaneously for both the material displacements (i.e. crack extension) and the spatial displacements, is adopted. In order to trace the dissipative loading path, an arc-length procedure is developed that controls the incremental crack area growth. In order to maintain mesh quality, smoothing of the mesh is undertaken as a continuous process, together with face flipping, node merging and edge splitting where necessary. Hierarchical basis functions of arbitrary polynomial order are adopted to increase the order of approximation without the need to change the finite element mesh. Performance of the formulation is demonstrated by means of three representative numerical simulations, demonstrating both accuracy and robustness.
\end{abstract}


\end{frontmatter}

%\linenumbers

\section{Introduction}

The pervasive and serious nature of cracks in materials and structures means that the computational modelling of crack propagation continues to be a critical area of research, and a major challenge. This paper presents a finite element based computational framework for modelling brittle crack propagation in elastic three-dimensional solids, based on the concept of configurational mechanics. The focus of this paper is on both the mathematical formulation and the computational framework to model and continuously resolve propagating cracks in a robust and computationally tractable manner. This paper is concerned with quasi-static problems where the influence of inertia is ignored. A sequel to this paper will demonstrate the extension of this work to dynamic fracture. 

%This paper describes both the mathematical framework and the computational implementation for crack propagation. 
%A new approach for predicting the direction and extent of the advancing crack front is presented. This is implemented into a FE framework where the crack is resolved as a displacement discontinuity, with the the front's motion modelled as a smooth and continuous process. 

The concept of configurational mechanics dates back to the original work of Eshelby and his study of forces acting on continuum defects~\cite{R32,R33}. More recently, configurational mechanics has been adopted by, amongst others, Maugin~\cite{R34, R35}, Steinmann~\cite{R36}, Miehe~\cite{R5,R6}.
In the context of this paper, a configurational change is exhibited as an advancing crack front. To formulate the crack propagation problem, two related kinematic descriptions are defined in the spatial and material settings. In the former, the classical conservation law of linear momentum balance is described, where Newtonian forces are work conjugate to changes in the spatial position, at fixed material position (i.e. no crack propagation). In the material setting, which represents a dual to the spatial setting, an equivalent conservation law is described, where configurational forces are conjugate to changes in material position but with no spatial motion. This decomposition of the behaviour is proven to be a simple but powerful methodology for describing crack propagation.

The spatial and material displacements fields are both discretised using the same finite element mesh, although we adopt different levels of approximation for the two fields. The resulting discretised weak form of the two conservation equations represent a set of coupled, nonlinear, algebraic equations that are solved in a monolithic manner using a Newton-Raphson scheme. In addition, an arc-length method, using crack area rather than displacements as a constraint, is adopted to trace the dissipative load path.

The resulting stress-free crack is represented as a displacement discontinuity, requiring the finite element mesh to resolve the crack and to be continuously adapted as the crack front advances. In contrast to other approaches, where the mesh is modified in order to resolve the incremental crack front advancement~\cite{R5,R6,R31}, in this paper we establish an equilibrium condition for the nodal configurational forces acting on the crack front that enables the crack front to advance in a continuous manner. In order to maintain mesh quality, smoothing of the mesh is undertaken as a continuous process, together with face flipping, node merging and edge splitting where necessary. With this methodology, it is important to note that there is no need to post-process stresses to determine if the crack should propagate and the crack front shape is calculated based purely on the physical equations. 
Three numerical examples are presented that demonstrate the ability of the formulation to accurately predict crack paths without bias from the original mesh.

\section{Body and crack kinematics}

\begin{figure}[th]
\setlength{\fboxsep}{0pt}%
\setlength{\fboxrule}{0pt}%
\begin{center}
\includegraphics[width=0.7\textwidth]{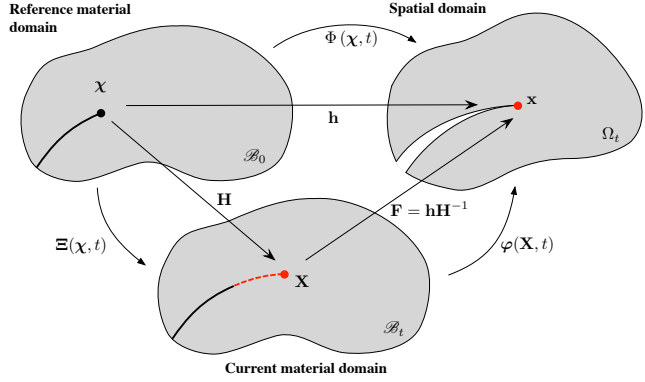}
\end{center}
\caption{Decomposition of crack propagation in elastically deforming body \label{F1}}
\end{figure}

Figure~\ref{F1} shows an elastic body with an initial crack in the reference material domain. As a result of loading, the crack extends and the body deforms elastically. It is convenient to decompose this behaviour into a purely configurational change, i.e. crack extension, which is described by the mapping from the reference material domain to the current material domain (${\boldsymbol\Xi}$), followed by elastic deformation only, described by the mapping from the current material to spatial domain (${\boldsymbol\varphi}$). We utilise these mappings to independently observe the evolution of the crack surface in the material domain  $\mathscr{B}_t$ and the elastic deformation of solid in the spatial domain $\Omega_t$.

The material coordinates $\mathbf{X}$ are mapped onto the spatial coordinates $\mathbf{x}$ via the
familiar deformation map $\boldsymbol\varphi(\mathbf{X},t)$. The physical displacement is:
\begin{equation}
\mathbf{u}=\mathbf{x}-\mathbf{X}
\end{equation}
The reference material domain describes the body before crack extension. ${\boldsymbol\Xi}(\boldsymbol\chi,t)$ maps the reference material coordinates $\boldsymbol\chi$ on to the current material coordinates $\mathbf{X}$, representing a configurational change, i.e. extension of the crack due to advancement of the crack front. ${\boldsymbol\Phi}$ maps the reference material coordinates $\boldsymbol\chi$ on to the spatial coordinates $\mathbf{x}$. The current material and spatial displacement fields are given as
\begin{equation}
\mathbf{W} = \mathbf{X} - {\boldsymbol\chi}\quad\textrm{and}\quad
\mathbf{w} = \mathbf{x} - {\boldsymbol\chi}
\end{equation}
$\mathbf{H}$ and $\mathbf{h}$ are the gradients of the material and spatial maps and $\mathbf{F}$ the deformation gradient~\cite{R31}, defined as:
\begin{equation}
\mathbf{H}=\frac{\partial {\boldsymbol\Xi}}{\partial {\boldsymbol\chi}},\quad\mathbf{h}=\frac{\partial {\boldsymbol\Phi}}{\partial {\boldsymbol\chi}},\quad\mathbf{F} = \frac{\partial \boldsymbol\varphi}{\partial \mathbf{X}} = \mathbf{h}\mathbf{H}^{-1}
\end{equation}

Given that the physical material cannot penetrate itself or reverse the movement of material coordinates, we have:
\begin{equation} \label{eq:det}
\textrm{det}(\mathbf{F}) = 
\frac{\textrm{det}(\mathbf{h})}
{\textrm{det}(\mathbf{H})} > 0
\end{equation}

In addition, the time derivative of the physical displacement $\mathbf{u}$ and the deformation gradient $\mathbf{F}$ are given as~\cite{R31}:
\begin{equation}\label{eq:phy_vel}
\dot{\mathbf{u}}= \dot{\mathbf{w}}-\mathbf{F}\dot{\mathbf{W}} \qquad 
\dot{\mathbf{F}} =
\nabla_{\mathbf{X}} \dot{\mathbf{w}} - \mathbf{h}\mathbf{H}^{-1}\nabla_{\mathbf{X}} \dot{\mathbf{W}}
\end{equation}

\section{Crack Description}

\begin{figure}[th]
\setlength{\fboxsep}{0pt}%
\setlength{\fboxrule}{0pt}%
\begin{center}
\includegraphics[width=0.8\textwidth]{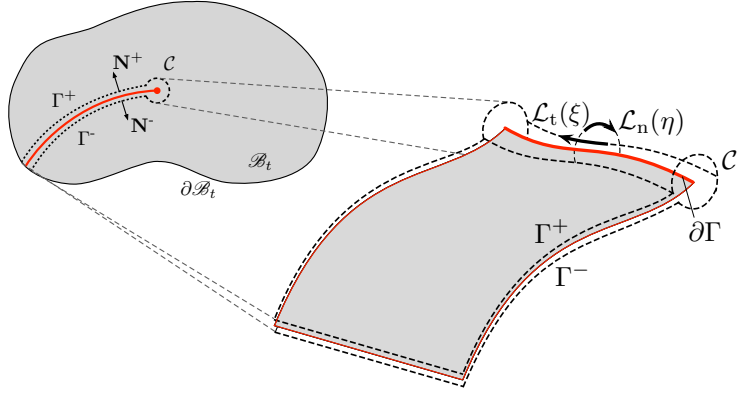}
\end{center}
\caption{Crack construction. In 2D (left) and in more detail in 3D (right).\label{F2}}
\end{figure}

The two faces of the crack surface $\Gamma$ are denoted by $\Gamma^+$ and $\Gamma^-$, the crack front is denoted as $\partial\Gamma$
and $\mathcal{C}$ is a surface that encircles the crack front, as shown in Figure~\ref{F2}. The crack surface can be described by local coordinates $\xi$ and $\eta$, such that $\Gamma =
\Gamma(\xi,\eta)$.
The area of the crack surface is given as:
\begin{equation}
A_\Gamma = 
\frac{1}{2}
(A^+_\Gamma + A^{-}_\Gamma) = 
\frac{1}{2}
\sum_{i}^{+,-}
\int_{\Gamma^i} \| \mathbf{N}^{i} \| \textrm{d}\xi \textrm{d}\eta
\end{equation}
where $\mathbf{N}^{+,-}$ is the normal to the crack faces $\Gamma^{+,-}$. The change of the crack surface area in the material domain is expressed as
\begin{equation} \label{eq::Agamma}
\begin{split}
\dot{A}_\Gamma &= 
\frac{1}{2} 
\sum^{+,-}_{i}
\int_{\Gamma^i}
\left\{
\frac{\mathbf{N}^{i}}{\| \mathbf{N}^{i} \|} 
\cdot
\left(
\textrm{Spin}\left[\mathbf{T}^{i}_\xi\right]
\frac{\partial\mathbf{T}^{i}_\eta}{\partial\mathbf{W}}
-
\textrm{Spin}\left[\mathbf{T}^{i}_\eta\right]
\frac{\partial\mathbf{T}^{i}_\xi}{\partial\mathbf{W}}
\right) 
\right\}\dot{\mathbf{W}}
\textrm{d}\xi \textrm{d}\eta\\
& = 
\frac{1}{2} 
\sum^{+,-}_{i}
\int_{\Gamma^i}
\mathbf{A}_{\Gamma^i} \cdot \dot{\mathbf{W}}
\textrm{d}S
\end{split}
\end{equation}
where $\mathbf{T}^{+,-}_\xi$ and $\mathbf{T}^{+,-}_\eta$ are the tangent vectors to
the crack faces $\Gamma^{+,-}$, and the crack face normals are defined as $\mathbf{N}^{+,-}
=\mathbf{\mathbf{T}^{+,-}_\xi} \times \mathbf{T}^{+,-}_\eta=\textrm{Spin}[\mathbf{T}^{+,-}_\xi]\mathbf{T}^{+,-}_\eta$.

It is convenient to parameterise the surface $\mathcal{C}$ using two families
of curves $\mathcal{L}_\textrm{t}(\xi)=\mathcal{C}(\xi,\eta)|_{\eta=\textrm{const}}$ and 
$\mathcal{L}_\textrm{n}(\eta)=\mathcal{C}(\xi,\eta)|_{\xi=\textrm{const}}$, as shown in Figure~\ref{F2}. 
In the limit, $| \mathcal{C} | \to 0$ and $\Gamma^+,\Gamma^- \to \Gamma$, resulting in a single crack surface $\Gamma$ with a crack front $\partial\Gamma$.
Thus, $\mathbf{A}_{\Gamma^{+}}, \mathbf{A}_{\Gamma^{-}} \to \mathbf{A}_{\Gamma}$, which describes the current orientation of the crack surface (dimensions of inverse length)
and is well defined for every point on the surface $\Gamma$, including the crack front $\partial\Gamma$. 

Recognising that any change in the crack surface area $\dot{A}_\Gamma$ in the material space can only occur due to 
motion of the crack front, the integral over the crack surface $\Gamma$ in Eq.(\ref{eq::Agamma}) can be 
expressed equivalently as an integral over the crack front $\partial\Gamma$.

\begin{equation}
\label{eq::Agamma2}
\dot{A}_\Gamma
 = 
\int_{\Gamma}
\mathbf{A}_{\Gamma} \cdot \dot{\mathbf{W}}
\textrm{d}S
= 
\int_{\partial\Gamma} 
\lim_{|\mathcal{L}_\textrm{n}|\to0}
\int_{\mathcal{L}_\textrm{n}} 
\mathbf{A}_\Gamma \cdot \dot{\mathbf{W}} \textrm{d}S 
=
\int_{\partial\Gamma}
\mathbf{A}_{\partial\Gamma} \cdot \dot{\mathbf{W}} \textrm{d}L
\end{equation}
This defines a kinematic relationship between the change in the
crack surface area $\dot{A}_\Gamma$ and the crack front velocity $\dot{\mathbf{W}}$
as an alternative to Eq.(\ref{eq::Agamma}), where 
$\mathbf{A}_{\partial\Gamma}$ is a dimensionless kinematic state variable that defines the current orientation of the crack front. 

\section{Dissipation of energy due to creation of new crack surfaces}

Consider an elastic body whereby energy dissipation is restricted to an increase in the crack surface area. Making use of Eq.(\ref{eq:phy_vel}), the power of
external work on the elastic body is given as: 
\begin{equation}
\mathscr{P} := \int_{\partial\mathscr{B}_t} \dot{\mathbf{u}} \cdot\mathbf{t} \textrm{d}S 
= \int_{\partial\mathscr{B}_t} \left\{\dot{\mathbf{w}}\cdot\mathbf{t}-\dot{\mathbf{W}}\cdot\mathbf{F}^\textrm{T}\mathbf{t}  \right\} dS
\end{equation} 
where $\mathbf{t}$ is the external traction vector. The rate of change of internal energy of the system can be decomposed as follows:
\begin{equation} \label{eq::it_work} \dot{\mathscr{U}} :=
\dot{\mathscr{U}}_{\Gamma} + \dot{\mathscr{U}}_{\mathscr{B}_t} 
\end{equation}
where $\mathscr{U}_{\Gamma}$ is the internal crack energy and $\mathscr{U}_{\mathscr{B}_t}$ is the internal body energy. The former is defined as:
\begin{equation} 
\mathscr{U}_{\Gamma} :=
%\frac{1}{2} \sum_{i}^{+,-} \gamma A_{\Gamma^i} 
\gamma A_{\Gamma} 
\end{equation} 
where $\gamma$ is the
surface energy and has dimension $[Nm^{-1}]$. Given Eq. (\ref{eq::Agamma2}), the change of the crack surface internal
energy is expressed as:
\begin{equation} 
\dot{\mathscr{U}}_{\Gamma} :=
\frac{\textrm{d}}{\textrm{d}t}\mathscr{U}_{\Gamma} = \gamma \dot{A}_\Gamma=
%\frac{1}{2} \sum_{i}^{+,-} \gamma \dot{A}_\Gamma^i =
%\int_{\partial\Gamma}  \gamma \mathbf{A}_{\partial\Gamma} \cdot \dot{\mathbf{W}} \textrm{d}L
\gamma \int_{\partial\Gamma}  \mathbf{A}_{\partial\Gamma} \cdot \dot{\mathbf{W}} \textrm{d}L
\end{equation} 
%The above establishes a link between the material displacements
%$\dot{\mathbf{W}}$ and the change in crack area $\dot{A}_\Gamma$. 
Furthermore, the change of
internal body energy is expressed as 
\begin{equation}
\label{int_en}
\dot{\mathscr{U}}_{\mathscr{B}_t} :=
\frac{\textrm{d}}{\textrm{d}t}\int_{\mathscr{B}_t} \Psi(\mathbf{F})
\textrm{d}V, 
\end{equation} 
where $\Psi$ is the specific free energy. Given the relation in Eq.(\ref{eq:phy_vel}) and that 
$\textrm{d}\dot{V} = \nabla_\mathbf{X} \cdot \dot{\mathbf{W}} \textrm{d}V$, Eq.(\ref{int_en}) can also be expressed as
\begin{equation} 
\dot{\mathscr{U}}_{\mathscr{B}_t} = 
\int_{\mathscr{B}_t} \left(
\mathbf{P}:\nabla_{\mathbf{X}}\dot{\mathbf{w}} +
{\boldsymbol\Sigma}:\nabla_{\mathbf{X}}\dot{\mathbf{W}} \right) \textrm{d}V 
\end{equation} 
where \begin{equation} 
\mathbf{P}:=\frac{\partial \Psi(\mathbf{F})}{\partial \mathbf{F}},\quad
{\boldsymbol\Sigma} :=
\Psi(\mathbf{F})\mathbf{1}-\mathbf{F}^\textrm{T}\mathbf{P}
\end{equation} 
are the first Piola-Kirchhoff stress and Eshelby stress tensors, respectively. The Piola-Kirchhoff stress is the familiar driving force for elastic deformation in the spatial domain, whereas the Eshelby stress is its material counterpart and the driving force for local configurational changes.
Thus, the first law of thermodynamics, $\mathscr{P}=\dot{\mathscr{U}}_{\Gamma} + \dot{\mathscr{U}}_{\mathscr{B}_t} $,  
can be expressed as
%which satisfy the
%polyconvexity requirement, therefore ensuring the existence of minimisers in
%variational problems appearing in the framework of the finite element method.
%Making use of Eqs.~(\ref{eq:phy_vel},\ref{eq::gard_vel}) and
%$\textrm{d}\dot{V} = \nabla_\mathbf{X} \cdot \dot{\mathbf{W}} \textrm{d}V$ in
%Eq.~(\ref{eq::it_work}), the first law is expressed by
\begin{equation} \label{eq::first_law} 
\int_{\partial\mathscr{B}_t} \left\{
\dot{\mathbf{w}}\cdot\mathbf{t}-\dot{\mathbf{W}}\cdot\mathbf{F}^\textrm{T}\mathbf{t}
\right\} \textrm{d}S 
=
\int_{\partial\Gamma} \gamma
\mathbf{A}_{\partial\Gamma} \cdot \dot{\mathbf{W}} \textrm{d}L + \int_{\mathscr{B}_t} \{
\mathbf{P}:\nabla_{\mathbf{X}}\dot{\mathbf{w}} +
{\boldsymbol\Sigma}:\nabla_{\mathbf{X}}\dot{\mathbf{W}} \} \textrm{d}V \end{equation} 
In order to get a
local form of the first law, the Gauss divergence theorem is applied to the last
integral in Eq.(\ref{eq::first_law}) resulting in the following expression 
\begin{equation} \label{eq::div_U} 
\begin{split} 
\int_{\partial\Gamma} \gamma \mathbf{A}_{\partial\Gamma}
\cdot \dot{\mathbf{W}} \textrm{d}L &=
 \int_{\mathscr{B}_t}
\dot{\mathbf{w}}\cdot\{ \nabla_\mathbf{X} \cdot \mathbf{P} \} \textrm{d}V +
\int_{\mathscr{B}_t} \dot{\mathbf{W}}\cdot\{ \nabla_\mathbf{X} \cdot
{\boldsymbol\Sigma} \} \textrm{d}V \\ &+
\int_{\partial\mathscr{B}_t\cup\Gamma^+\cup\Gamma^-}
\dot{\mathbf{w}}\cdot\{\mathbf{t} - \mathbf{P}\mathbf{N} \} \textrm{d}S +
\int_{\partial\mathscr{B}_t\cup\Gamma^+\cup\Gamma^-}
\dot{\mathbf{W}}\cdot\{\mathbf{F}^\textrm{T}\mathbf{t}+\boldsymbol\Sigma\mathbf{N}\}
\textrm{d}S \\ &- 
\lim_{|\mathcal{C}|\to0} \int_{\mathcal{C}} 
\dot{\mathbf{w}} \cdot
\mathbf{P}\mathbf{N} \textrm{d}S + 
\lim_{|\mathcal{C}|\to0} \int_{\mathcal{C}} 
\dot{\mathbf{W}} \cdot
\boldsymbol\Sigma
\mathbf{N}\textrm{d}S
\end{split} 
\end{equation} 
To simplify this equation, we recognise that, in the limit, the surface $\mathcal{C}$ collapses to the crack front $\partial\Gamma$, and integrals over the crack front can be expressed as
\begin{equation}
\lim_{|\mathcal{C}|\to0} 
\int_{\mathcal{C}} (\cdot) \textrm{d}S := 
\lim_{|\mathcal{C}|\to0} 
\int_{\mathcal{L}_\textrm{t}} \int_{\mathcal{L}_\textrm{n}} (\cdot) \textrm{d}S
=
\int_{\partial\Gamma} \lim_{|\mathcal{L}_\textrm{n}|\to0} \int_{\mathcal{L}_\textrm{n}} (\cdot) \textrm{d}S
\end{equation}
In addition, the first and second terms on the right hand side of Eq.(\ref{eq::div_U}) vanish since the the spatial and material conservation laws of linear momentum balance are expressed as follows:
\begin{equation}
\nabla_\mathbf{X} \cdot \mathbf{P} =\mathbf{0},\quad\nabla_\mathbf{X} \cdot
{\boldsymbol\Sigma} =\mathbf{0}
\end{equation}
Moreover, considering only admissible velocity fields and stress fields in
equilibrium with external forces, Eq.(\ref{eq::div_U}) can be expressed as:
\begin{equation}
\int_{\partial\Gamma} \gamma \mathbf{A}_{\partial\Gamma} \cdot \dot{\mathbf{W}}\,
\textrm{d}L
-
\int_{\partial\Gamma} \dot{\mathbf{W}} \cdot 
\lim_{|\mathcal{L}_\textrm{n}|\to0} \int_{\mathcal{L}_\textrm{n}}
\boldsymbol\Sigma\mathbf{N} \, \textrm{d}S = 0 \end{equation}
Defining the configurational force as
\begin{equation}
\mathbf{G}=\lim_{|\mathcal{L}_\textrm{n}|\to0} \int_{\mathcal{L}_\textrm{n}}
\boldsymbol\Sigma\mathbf{N} \; \textrm{d}L
\end{equation}
the local form of the first law (Eq.(\ref{eq::div_U})) is expressed as:
\begin{equation} \label{eq::first_law_local} 
\dot{\mathbf{W}} \cdot \left( \gamma \mathbf{A}_{\partial\Gamma}
-
\mathbf{G} \right) 
= 0
\end{equation} 
This equation represents the equilibrium condition for the crack front. The term $\gamma \mathbf{A}_{\partial\Gamma}$ can be considered the material resistance and, in the case of the crack front evolving, we can deduce that $\dot{\mathbf{W}}\ne 0$ and $\gamma \mathbf{A}_{\partial\Gamma}=\mathbf{G}$. 
It should be noted that for a continuous elastic body comprising an homogeneous material, the configurational forces $\mathbf{G}$ within the volume, away from the crack front, should be zero. 

Since all energy dissipation is restricted to creation of new crack surfaces, it follows that the local form of the second law is given as
\begin{equation} \label{eq::second_law} 
\mathscr{D} := \gamma
\dot{\mathbf{W}} \cdot \mathbf{A}_{\partial\Gamma} = \dot{\mathbf{W}} \cdot \mathbf{G} \geq 0
\end{equation}
where $\mathscr{D}$ is the dissipation of energy per unit length of the crack front. This inequality restricts evolution of the crack to positive crack area growth at each point of
the crack front. Although the first law defines if the crack front is in equilibrium and the second law places restrictions on the
direction of crack evolution, it does not determine how $\mathbf{A}_{\partial\Gamma}$ or
$\dot{\mathbf{W}}$ evolves. In the next section we supplement the second law with a crack growth criterion. 

\section{Evolution of the crack front}

A straightforward criterion for crack growth, in the spirit of Griffith, is proposed:
\begin{equation} \label{eq:grif1}
\phi(\mathbf{G}) = 
\mathbf{G} \cdot \mathbf{A}_{\partial\Gamma} - g_c/2 \leq 0
\end{equation} 
where $g_c=2\gamma$ is a material parameter specifying the critical threshold of energy release
per unit area of the crack surface $\Gamma$. For a point on the crack front to be in equilibrium, 
either $\phi<0$ and $\dot{\mathbf{W}}=0$, or $\phi=0$, $\dot{\mathbf{W}}\ne 0$ and $\gamma\mathbf{A}_{\partial\Gamma}=\mathbf{G}$ must be satisfied. 

To determine the evolution of a point on the crack front we adopt the principle of maximum dissipation, that states
that, for all possible configurational forces $\mathbf{G}^*$
satisfying the crack growth criterion $\phi(\mathbf{G}^*) = 0$, the
dissipation $\mathscr{D}$ attains its maximum for the actual configurational  
force $\mathbf{G}$. Therefore, we have
\begin{equation}
\mathscr{D}_\textrm{max}=(\mathbf{G} - \mathbf{G}^*)\cdot\dot{\mathbf{W}} \geq 0
\end{equation}
This can be interpreted as a unconstrained minimisation problem, for which the Lagrangian function is:
\begin{equation}
\mathscr{L}(\mathbf{G}^*,\dot{\kappa}) = - \mathbf{G}^*\cdot\dot{\mathbf{W}} + \dot{\kappa}\phi(\mathbf{G}^*)
\end{equation}
with $\dot{\kappa}$ the Lagrange multiplier. The Kuhn-Tucker conditions become
\begin{equation}
\frac{\partial\mathscr{L}}{\partial\mathbf{G}^*}=-\dot{\mathbf{W}}+\dot{\kappa}\mathbf{A}_{\partial\Gamma}=0,\quad
\dot{\kappa} \geq 0,\quad\phi(\mathbf{G}^*_{\partial\Gamma}) \leq 0,\quad\textrm{and}\quad\dot{\kappa}\phi(\mathbf{G}^*) = 0
\end{equation}
Thus, for a point on the evolving crack front, the crack front orientation is colinear to the configurational force (Eq.(\ref{eq::first_law_local})), i.e. $\gamma \mathbf{A}_{\partial\Gamma}=\mathbf{G}$, and the crack extension is given as $\dot{\mathbf{W}}=\dot{\kappa}\mathbf{A}_{\partial\Gamma}$.
%
%
%
%In an unloading situation, i.e. $\phi<0$, the trivial solution of no crack growth is obrained, i.e.
%\begin{equation} \label{eq::front0}
%\dot{\mathbf{W}} = \mathbf{0},
%\end{equation}
%where the first law (\ref{eq::first_law}) is automatically satisfied and the crack
%orientation $\mathbf{A}_{\partial\Gamma}$ is determined by the current crack geometry. For loading 
%of a point on the crack front, $\dot{\kappa} > 0$ (and $\phi = 0$),
%\begin{equation} \label{eq::front1}
%\dot{\mathbf{W}} = \dot{\kappa}\mathbf{A}_{\partial\Gamma}
%\end{equation}
%which constrains the crack extension $\dot{\mathbf{W}}$ to be collinear to
%$\mathbf{A}_{\partial\Gamma}$. This result identifies the second possible solution to the first law is ... i.e.
%\begin{equation} \label{eq::front2}
%\gamma \mathbf{A}_{\partial\Gamma} = \mathbf{G}_{\partial\Gamma} \quad \textrm{and} \quad 2 \gamma = g_c
%\end{equation}
%Thus, applying the principle of maximum dissipation, we determine that the crack propagation direction is determined by the configurational force.
%%A thermodynamically admissible crack propagation is only possible for a
%%positive local change of the crack surface area (see
%%Eq.~(\ref{eq::second_law}). 
$\dot{\kappa}$ has the dimension of length
and represents the kinematic state variable for a point on the crack front, which can be
identified as:
\begin{equation}
\dot{A}_\Gamma \equiv \int_{\partial\Gamma}\dot{\kappa}\,\textrm{d}L \geq 0
\end{equation}

\section{Spatial and material discretisation}

Finite element approximation is applied for displacements in both the current material and physical space
\begin{equation}
\begin{array}{c}
\mathbf{X}^\textrm{h} = {\boldsymbol\Phi}(\boldsymbol\chi)\widetilde{\mathbf{X}},\quad
\mathbf{x}^\textrm{h} = {\boldsymbol\Phi}(\boldsymbol\chi)\widetilde{\mathbf{x}}\\
\mathbf{W}^\textrm{h} = {\boldsymbol\Phi}(\boldsymbol\chi)\dot{\widetilde{\mathbf{W}}},\quad
\mathbf{w}^\textrm{h} = {\boldsymbol\Phi}(\boldsymbol\chi)\dot{\widetilde{\mathbf{w}}}
\end{array}
\end{equation}
where superscript $\textrm{h}$ indicates approximation and $\left(\tilde{\cdot}\right)$ indicates nodal values. Three-dimensional domains are discretised with tetrahedral finite elements. In the spatial domain, hierarchical basis functions of arbitrary polynomial order are applied, following the work of Ainsworth and Coyle~\cite{R12}. This enables the use of elements of variable, non-uniform order approximation, with conformity enforced across element boundaries. In the material domain, linear approximation is adopted, as this is sufficient for describing the crack front. 

The discretised gradients of deformation are expressed as
\begin{equation}
\mathbf{H}^\textrm{h} = \mathbf{B}_\mathbf{X}(\boldsymbol\chi)\widetilde{\mathbf{X}},\quad
\mathbf{h}^\textrm{h} = \mathbf{B}_\mathbf{x}(\boldsymbol\chi)\widetilde{\mathbf{x}}, \quad
\mathbf{F^\textrm{h}} = \mathbf{h}^\textrm{h}(\mathbf{H}^\textrm{h})^{-1} = 
\mathbf{B}_\mathbf{x}(\mathbf{X})\widetilde{\mathbf{x}}
\end{equation}
%and the physical deformation gradient can be expressed as:
%\begin{equation}
%\mathbf{F^\textrm{h}} = \mathbf{h}^\textrm{h}(\mathbf{H}^\textrm{h})^{-1} = 
%\mathbf{B}_\mathbf{x}(\mathbf{X})\widetilde{\mathbf{x}}
%\end{equation}

\subsection{Crack orientation}

The normal to the discretised crack surface $\Gamma^\textrm{h}$, applying the FE approximation, is given by:
\begin{equation}
\label{eq::crack_normal_approx}
\mathbf{N}^\textrm{h} =
\textrm{Spin}\left[
\frac{\partial\mathbf{X}^\textrm{h}}{\partial \xi}
\right]
\frac{\partial\mathbf{X}^\textrm{h}}{\partial \eta} 
\end{equation}
where $\xi$ and $\eta$ are local coordinates of an element's triangular face on the crack surface.
This normal is constant for a linear element and is easily calculated at 
Gauss integration points for
higher-order approximations. Utilising Eq.(\ref{eq::crack_normal_approx}), and with reference to Eq.(\ref{eq::Agamma}), the approximation to the change in crack area can be expressed as
\begin{equation}
\begin{split}
\dot{A}_\Gamma^\textrm{h} &=\frac{1}{2}
\left\{
\A_\textrm{TRI} \int_{\textrm{TRI}}
\frac{\mathbf{N}}{\| \mathbf{N} \|} 
\cdot
\left(
\textrm{Spin}\left[\frac{\partial \mathbf{X}^\textrm{h}}{\partial \xi}\right]
\frac{\partial {\boldsymbol\Phi}}{\partial \mathbf{\eta}}
-
\textrm{Spin}\left[\frac{\partial \mathbf{X}^\textrm{h}}{\partial \mathbf{\eta}}\right]
\frac{\partial {\boldsymbol\Phi}}{\partial \xi} 
\right) 
\textrm{d}\xi \textrm{d}\eta
\right\}
\dot{\widetilde{\mathbf{W}}} \\
&= \frac{1}{2}
\left\{
\A_\textrm{TRI} \int_{\textrm{TRI}}
\mathbf{A}^\textrm{h}_\Gamma
\textrm{d}\xi \textrm{d}\eta
\right\}
\dot{\widetilde{\mathbf{W}}} =
\widetilde{\mathbf{A}}^\textrm{h}_\Gamma 
\dot{\widetilde{\mathbf{W}}}
\end{split}
\end{equation}
The matrix
$\widetilde{\mathbf{A}}^\textrm{h}_\Gamma$ has dimensions of length and describes the current orientation of the crack surface. Admissible changes in material nodal coordinates are restricted to nodes on the crack front $\partial\Gamma^\textrm{h}$.
%$I \in \{I: \mathcal{N}_I\,\textrm{is node on the crack front } \partial\Gamma^\textrm{h}\}$. 
As a consequence, the number of non-zero rows of the matrix $\widetilde{\mathbf{A}}^\textrm{h}_\Gamma$ is equal to the number of active crack front nodes. 
$\A_\textrm{TRI}$ indicates the standard FE assembly for
the triangular faces of 3D tetrahedral elements elements that lie on the crack surface.

%Each column of $\widetilde{\mathbf{A}}^\textrm{h}_\Gamma$ represents the contribution of the unit nodal material velocities $\dot{\widetilde{\mathbf{W}}}$ to changes of the crack surface area $A^\textrm{h}_\Gamma$.  
%Each row of matrix $\widetilde{\mathbf{A}}^\textrm{h}_\Gamma$ can be interpreted as the components of
%the corresponding node's material velocity which lead to a change of crack area. 

\subsection{Residuals in spatial and material domain}

The residual force vector in the discretised spatial domain is expressed in the classical way as:
\begin{equation}
\label{eq::spa_residual}
\mathbf{r}^\textrm{h}_\textrm{s} := 
\tau\mathbf{f}^\textrm{h}_\textrm{ext,s}
-\mathbf{f}^\textrm{h}_\textrm{int,s} 
%= \mathbf{0}
\end{equation}
where $\tau$ is the unknown scalar load factor and $\mathbf{f}^\textrm{h}_\textrm{ext,s}$ and $\mathbf{f}^\textrm{h}_\textrm{int,s}$ are the vectors of external and internal forces, respectively, and defined as follows
\begin{equation}
\mathbf{f}^\textrm{h}_\textrm{ext,s} := 
\A_\textrm{TRI} \int_{\textrm{TRI}} {\boldsymbol\Phi}^\textrm{T}\mathbf{t}^\textrm{h}_\tau \textrm{d}S,\quad
\mathbf{f}^\textrm{h}_\textrm{int,s} = 
\A_\textrm{TET} \int_{\textrm{TET}} \left(\mathbf{B}_\mathbf{x}\right)^\textrm{T} \mathbf{P}^\textrm{h} \textrm{d}S
\end{equation}
where $\A_\textrm{TET}$ indicates the standard FE assembly for tetrahedral elements.  

The residual vector in the material domain, as a counterpart to Eq.(\ref{eq::spa_residual}), is given by
\begin{equation}
\label{eq::material_residual}
\mathbf{r}^\textrm{h}_\textrm{m} := 
\mathbf{f}^\textrm{h}_\textrm{res}
-\widetilde{\mathbf{G}}^\textrm{h}
% +\tilde{\mathbf{f}}^\textrm{h}_\textrm{quality} 
%= \mathbf{0}
\end{equation}
$\mathbf{f}^\textrm{h}_\textrm{res}$ is the vector of nodal material resistance forces, given as:
\begin{equation}
\label{eq::mat_res}
\mathbf{f}^\textrm{h}_\textrm{res} := 
\frac{1}{2}\left(
\widetilde{\mathbf{A}}^\textrm{h}_\Gamma 
\right)^\textrm{T}
\mathbf{g}_c
\end{equation}
where $\mathbf{g}_c$ is a vector of size equal to the number of nodes in the mesh, with zero for all components except those associated with nodes on the crack front, where the value is $g_c$. These nodal forces of material resistance have dimensions of force and are work conjugate to the material displacement $\mathbf{W}$ on the crack front. $\widetilde{\mathbf{G}}^\textrm{h}$ are the nodal configurational forces, and considered the driving force for crack front evolution.
\begin{equation} \label{eq::nodal_resistance}
\widetilde{\mathbf{G}}^\textrm{h}:=
\A_\textrm{TET} \int_{\textrm{TET}} \left(\mathbf{B}_\mathbf{x}\right)^\textrm{T} {\boldsymbol\Sigma}^\textrm{h} \textrm{d}S
\end{equation}

The nodal configurational forces are calculated and used only for nodes on the crack front. Non-zero values in the rest of the domain are negligible compared to these. However, it is recognised that non-zero configurational forces away from the crack front can indicate a discretisation error and this has been utilised in other work as a driver for solution improvement~\cite{R13} using r-adaptivity. In this work, in the vicinity of the crack front, the order of approximation and mesh density are increased when the magnitude of configurational forces away from the crack front are numerically significant.

\subsection{Arc-length control} \label{arc}
The global equilibrium solution for the spatial and material displacements is obtained as a fully coupled problem using the Newton-Raphson method, converging when the norms of the residuals are less than a given tolerance. 
To trace the nonlinear response resulting from the dissipative behaviour, an arc-length technique is adopted. 
Thus, the system of equations for conservation of the material
and spatial momentum is supplemented by a load control equation that imposes a constraint that conserves the total crack area during each load step. This load control equation takes the form:
\begin{equation} \label{eq::control}
r_\tau = \sum_\textrm{I} (\widetilde{\mathbf{A}}^\textrm{h}_\Gamma \widetilde{\mathbf{X}}_n)_\textrm{I}-
\sum_\textrm{I}
(\widetilde{\mathbf{A}}^\textrm{h}_\Gamma \widetilde{\mathbf{X}}^{i+1}_{n+1})_\textrm{I} = 0,
\quad\textrm{for}\quad I \in \{I: \mathcal{N}_I\,\textrm{is crack front node}\}
\end{equation}
where $i+1$ is the current iteration of load step $n+1$.  

\section{Resolution of the Propagating Crack and Mesh Quality Control}
In the previous work of the authors and others \cite{R5,R6,R31}, a discrete face-splitting methodology was adopted. In that approach, the new crack front was generated by first identifying element faces ahead of the current crack front, then aligning them to the direction of the configurational forces and finally splitting these faces if the crack criterion was violated, creating a displacement discontinuity. This process was continued for all nodes on the crack front until equilibrium was achieved. In this paper, an alternative approach is proposed, whereby the crack front evolves in an implicit, continuous manner. The mesh is subsequently moved to resolve the new crack geometry (rather than changing the mesh to create the new crack front). At the end of each load step, equilibrium has been achieved and the configurational forces are parallel to the vector $\mathbf{A}_{\partial\Gamma}$. 

In the process of moving the mesh to resolve the moving crack front, the mesh can become distorted, potentially creating poor quality elements leading to numerical errors. To mitigate this effect we adopt several strategies:
\begin{enumerate}
\item  Edge splitting is applied to elements behind the crack front that have become too elongated. This action will result in the creation of new elements. It is enforced if, for a given node, there exists an adjacent edge with a length greater than $1.5$ times the average edge length of all adjacent edges to the node.

\item Node merging is applied to elements ahead of the crack front that have become too contracted. This action will result in the removal of elements. It is enforced if, for a given node, there exists an adjacent edge with a length less than $1/3$ of the length of the longest edge adjacent to the node.

\item Face flipping is applied to elements in the vicinity of the crack front to ensure that a 3D Delaunay triangulation exists, with optimal internal angles. This is described in more detail in section~\ref{face_flip} below.
\end{enumerate}

These procedures are utilised, if necessary, at the beginning of each load step, before the Newton-Raphson iterations begin, when the solution is already out of equilibrium. Furthermore, in the case when new nodes are added, variables are transferred to the new mesh based upon approximation of the variables using the old mesh.

\begin{figure}[th]
\setlength{\fboxsep}{0pt}%
\setlength{\fboxrule}{0pt}%
\begin{center}
\includegraphics[width=1.0\textwidth]{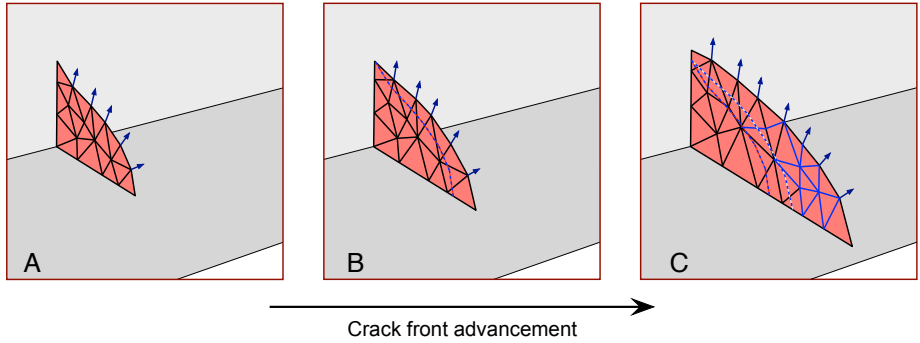}
\end{center}
\caption{Crack front advancement demonstrated with three-point bending of beam with initial edge notch. The lower images are snapshots of the propagating crack front, with the arrows representing the nodal configurational forces. The projection of the mesh on the crack faces is also shown, including new elements (shown in blue). The dotted lines indicate the position of the crack front in the previous snapshots.  \label{CF1}}
\end{figure}

Figure~\ref{CF1} demonstrates how the crack front evolves using the example of a three-point bending of a beam with an initial edge notch. Crack surface A is an equilibrium solution and the projection of the FE mesh onto the crack surface is shown. Also shown are the configurational forces. Crack surface B shows a subsequent configuration, where the crack front has advanced to a new equilibrium position. The mesh topology has remained the same but the nodes have moved to resolve this new crack geometry.
Crack surface C represents a further equilibrium configuration. Here it is clear that the mesh has changed, with new elements being created due to the edge splitting procedure behind the crack front.

\subsection{Face flipping}\label{face_flip}
\begin{figure}[th]
\setlength{\fboxsep}{0pt}%
\setlength{\fboxrule}{0pt}%
\begin{center}
\includegraphics[width=0.7\textwidth]{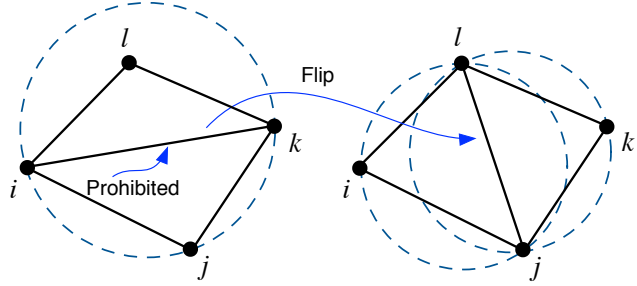}
\end{center}
\caption{Edge flipping in 2D to achieve Delaunay triangulation (blue dotted lines indicate circle that intersect nodes of corresponding element) \label{flip1}}
\end{figure}

At the beginning of each load step, when the solution is out of equilibrium, a patch of elements around the crack front is checked to ensure that it represents a 3D Delaunay triangulation. Figure~\ref{flip1} demonstrates the idea in 2D. Considering the two elements on the left, edge $i-k$ is prohibited because it lies in the interior of the circle that intersects the nodes of element $i-j-k$ (and element $i-k-l$). Flipping edge $i-k$ will address this problem, redefining the two adjacent elements, without affecting the rest of the mesh. Thus, edge $i-k$ is removed and replaced by edge $j-l$, and two new adjacent elements are formed that represent a Delaunay triangulation.

\subsection{Mesh quality control}
\label{sec::mesh_quality}

In addition to the mesh adjustments described above, a global mesh quality control procedure is also adopted. During the global Newton-Raphson procedure, constraints are imposed on the shape of element to ensure good mesh quality, but without influencing the physical response. 

The authors have proposed a volume-length measure of element quality~\cite{R31,R24}. Although this measure does not
directly determine dihedral angles, it has been shown to be very effective at
eliminating poor angles, thus improving stiffness matrix conditioning and reducing
interpolation errors \cite{R15,R16}. As the volume-length measure is a smooth
function of node positions, and its gradient is straightforward and
computationally inexpensive to calculate, it is ideal for the problem at hand. The
volume-length quality measure is defined as
\begin{equation} \label{eq::quality_mm}
q(\mathbf{H}^\textrm{h}) := 6\sqrt{2} \frac{V_0}{l_{\textrm{rms},0}^3}\frac{\textrm{det}(\mathbf{H}^\textrm{h})}{dl_{\textrm{rms}}^3(\mathbf{H}^\textrm{h})} 
= q_0b(\mathbf{H}^\textrm{h})
,\quad
b(\mathbf{H}^\textrm{h}) := \frac{\textrm{det}(\mathbf{H}^\textrm{h})}{dl_\textrm{rms}^3(\mathbf{H}^\textrm{h})}
\end{equation}
where $q_0$, $V_0$ and $l_{\textrm{rms},0}$ are the element quality, element volume and root mean square of the element's edge lengths respectively, in the reference configuration. $\mathbf{H}^\textrm{h}$ is the material deformation gradient, $b$ is a measure of element quality change, relative to the reference configuration, and $dl_{\textrm{rms}}=l_\textrm{rms}/l_{\textrm{rms},0}$ is the stretch of $l_{\textrm{rms},0}$. Element quality $q_0$ is normalised so that an equilateral element has quality $q_0=1$ and a degenerate element (zero volume) has
$q_0=0$. Furthermore, $b=1$ corresponds to no change and $b=0$ is a change leading to a degenerate element. An element edge length in the current material configuration is expressed as
\begin{equation}
l_j(\mathbf{H}^\textrm{h}) := \sqrt{
\Delta {\boldsymbol\chi}_j^\textrm{T}(\mathbf{H}^\textrm{h})^\textrm{T}\mathbf{H}^\textrm{h}\Delta {\boldsymbol\chi}_j
}
\end{equation}
where $\Delta {\boldsymbol\chi}^j$ is the distance vector of edge $j$ in the
reference configuration. Thus $l_\textrm{rms}$ is calculated as
\begin{equation}
l_\textrm{rms} :=
\sqrt{\frac{1}{6}
\sum_{j=1}^{6} l_j^2 }
 =  l_{\textrm{rms},0}\,dl_\textrm{rms}
\end{equation}
To control the quality of elements, an admissible deformation $\mathbf{H}^\textrm{h}$ is enforced
such that
\begin{equation} \label{eq::quali_const}
b(\mathbf{H}^\textrm{h})>\gamma \quad \textrm{for}\; \gamma \in (0,1)
\end{equation}
In practice, $0.1 < \gamma <
0.5$ gives good results. This constraint on $b$ is enforced  by applying a volume--length log--barrier function \cite{R13} defined, for the entire mesh, as
\begin{equation} \label{eq::mesh_energy}
%\mathcal{B} := \sum_{e=0}^\mathcal{N} \frac{q^2}{2(1-\gamma)} - \log(q_e-\gamma),\quad
\mathcal{B} := \sum_{e=0}^{\mathcal{N}_\textrm{el}} \frac{b^2}{2(1-\gamma)} - \ln(b-\gamma)
\end{equation}
where $\mathcal{B}$ is the barrier function for the change in element quality in the current material configuration and $\mathcal{N}_\textrm{el}$ is the number of elements. It can be seen that the log--barrier function
rapidly increases as the quality of an element reduces, and tends to infinity
when the quality approaches the barrier $\gamma$, thus achieving the objective of penalising the
worst quality elements. 

In order to build a solution scheme that incorporates a stabilising force that controls element quality, a pseudo `stress' at the element level is defined as a counterpart
to the first Piola--Kirchhoff stress as follows
\begin{equation} 
\mathbf{Q} :=  
\frac{\partial \mathcal{B} }{\partial \mathbf{H}^\textrm{h} } = 
\textrm{det}(\mathbf{H}^\textrm{h})
\frac{l_{\textrm{lrms},0}^3}{l_{\textrm{lrms}}^3}
 \left(
\frac{b}{1-\gamma} + \frac{1}{b-\gamma}
\right) \widehat{\mathbf{Q}},
\end{equation}
where matrix
$\widehat{\mathbf{Q}}$ is defined as follows
\begin{equation}\label{eq::quali_ene}
\widehat{\mathbf{Q}} := 
(\mathbf{H}^\textrm{h})^{-\textrm{T}}
-
\frac{1}{2}
\frac{1}{l_{\textrm{lrms}}^2}
\sum_{j=1}^6 
\Delta {\mathbf{X}}^j
(\Delta {\boldsymbol\chi}^j)^\textrm{T}.
\end{equation}
It is worth noting that $\mathbf{Q}$ should be a zero matrix for a purely volumetric change or rigid body movement of
a tetrahedral element. 

It is now possible to compute a vector of nodal pseudo `forces' associated with $\mathbf{Q}$ as
\begin{equation} \label{eq::rhs_mesh}
\mathbf{f}^\textrm{h}_\textrm{q} =
\A_\textrm{TET} 
\int_{\textrm{TET}} 
\mathbf{B}^\textrm{T}_\mathbf{X} \mathbf{Q}\,
\textrm{d}V 
\end{equation} 

\subsection{Shape preserving constraints}
\label{shape_constraints}
The continuous adaption of the mesh must be constrained to preserve the surface of the domain being analysed, including the crack surface. Nodes on the surface can only be allowed to slide along the surface and not deviate from it. In this work we start from the observation that a body's shape can be globally characterised as a constant:
\begin{equation}\label{eq:glob_shape_preserving_const}
\frac{V}{A}=C
\end{equation}
where $V$ is the volume of the body and $A$ is its surface area. This can be expressed in integral form as
\begin{equation}
\int_{\mathscr{B}_0} \textrm{d}V = C \int_{\partial\mathscr{B}_0} \textrm{d}S
\end{equation}
noting that 
\begin{equation}
\frac{1}{3} \int_{\mathscr{B}_0} \nabla \cdot \mathbf{X}_\textrm{s}\, \textrm{d}V = C \int_{\partial\mathscr{B}_0} \textrm{d}S
\end{equation}
where $\mathbf{X}_\textrm{s}$ are the coordinates on the surface $S$ in the current material domain. Applying Gauss theorem we obtain
\begin{equation}
\int_{\partial\mathscr{B}_0} \mathbf{X}_\textrm{s}\cdot \frac{\mathbf{N}}{\|\mathbf{N}\|}\, \textrm{d}S = 3C \int_{\partial\mathscr{B}_0} \textrm{d}S
\end{equation}
where $\mathbf{N}$ is the outer normal to the surface. The local form of this equation is given as
%Dropping integrals on both sides of the equation 
\begin{equation} \label{eq:51Sc}
\frac{\mathbf{N}}{\|\mathbf{N}\|}\cdot \mathbf{X}_\textrm{s}-3C=0\quad\mathbf{X} \in \partial\mathscr{B}_0
\end{equation}
where $3C$ is a constant given for the reference geometry. If the above constraint is satisfied, the body shape and volume are conserved locally. An equivalent form this constraint equation is
\begin{equation}
\frac{\mathbf{N}}{\|\mathbf{N}\|}\cdot (\mathbf{X}_\textrm{s}-\boldsymbol{\chi}_{\textrm{s}})=0\quad\mathbf{X} \in \partial\mathscr{B}_0
\end{equation}
where $\boldsymbol{\chi}_{\textrm{s}}$ are the original positions of the surface. 
%Finally we get a local form of equation (\ref{eq:glob_shape_preserving_const}) nonlinear equation on body surface. 

To enforce the mesh quality control described in the preceeding section, subject to the surface constraints described in this section,  the method of Lagrange multipliers is used, with the following functional,
\begin{equation}
\mathcal{L}(\widetilde{\mathbf{X}},\widetilde{\boldsymbol\lambda})=
\int_{\mathscr{B}_0} \mathbf{H}^\textrm{h}:\mathbf{Q}\, \textrm{d}V+
\widetilde{\boldsymbol\lambda}^\textrm{T}
\int_{\partial\mathscr{B}_0} \Phi^\mathsf{T}_\lambda \mathbf{N} \cdot \Phi_\mathbf{X} \, \textrm{d}\xi\textrm{d}\eta
\,
(\widetilde{\mathbf{X}}_\textrm{s}-\widetilde{\boldsymbol{\chi}}_{\textrm{s}}) 
\end{equation}
where $\mathbf{H}^\textrm{h}$ and $\mathbf{Q}$ are functions of material positions, $\widetilde{\mathbf{X}}$ and are current mesh nodal positions defined in the previous sections. $\xi$ and $\eta$ are parametrization of surface $\partial\mathscr{B}_0$. ${\boldsymbol\Phi}_\lambda$ and ${\boldsymbol\Phi}_\mathbf{X}$ are the shape functions for Lagrange multipliers and material coordinates respectively. ${\boldsymbol\Phi}_\lambda$ are piecewise continuous functions with order of approximation equal to that of the material coordinates.  

Calculating the stationary values of the Lagrangian results in the first $\left(\mathcal{L}(\widetilde{\mathbf{X}},\widetilde{\boldsymbol\lambda})_{,\widetilde{\mathbf{X}}}=0\right)$ and second $\left(\mathcal{L}(\widetilde{\mathbf{X}},\widetilde{\boldsymbol\lambda})_{,\widetilde{\boldsymbol\lambda}}=0\right)$ Euler equations. The first is given as 
\begin{equation} \label{eq:eq:glob_shape_preserving_const_weighted}
\int_{\partial\mathscr{B}_0} \Phi^\mathsf{T}_\lambda 
\mathbf{N} \cdot \Phi_\mathbf{X} \, \textrm{d}\xi\textrm{d}\eta\, 
(\widetilde{\mathbf{X}}_\textrm{s}-\widetilde{\boldsymbol\chi}_{\textrm{s},0}) 
= \mathbf{0}.
\end{equation}
Taking a truncated Taylor series expansion after the linear term of this nonlinear equation results in 
\begin{equation}
\label{euler1}
\begin{split} &
\int_{\partial\mathscr{B}_0} \Phi^\mathsf{T}_\lambda \left\{ \mathbf{N}\Phi_\mathbf{X} + \left(\mathbf{X}_\textrm{s}-{\boldsymbol\chi}_\textrm{s}\right) \cdot \left( \textrm{Spin}\left[\frac{\partial\mathbf{X}_\textrm{s}}{\partial\xi}\right]\cdot\mathbf{B}_\eta - \textrm{Spin}\left[\frac{\partial\mathbf{X}_\textrm{s}}{\partial\eta}\right]\cdot\mathbf{B}_\xi \right) \right\} \textrm{d}\xi\textrm{d}\eta \, \delta \widetilde{\mathbf{X}}_\textrm{s}\\ = &\int_{\partial\mathscr{B}_0} \Phi^\mathsf{T}_\lambda \mathbf{N}\cdot(\mathbf{X}_\textrm{s}-{\boldsymbol\chi}_{\textrm{s}})\, 
\textrm{d}\xi\textrm{d}\eta \end{split}.
\end{equation}
where the differential operators are defined as
\begin{equation}
\mathbf{B}_\xi\delta\mathbf{X}_\textrm{s}=\textrm{Spin}\left[\frac{\partial\delta\mathbf{X}_\textrm{s}}{\partial\xi}\right]
\quad\textrm{and}\quad
\mathbf{B}_\eta\delta\mathbf{X}_\textrm{s}=\textrm{Spin}\left[\frac{\partial\delta\mathbf{X}_\textrm{s}}{\partial\eta}\right].
\end{equation}
Linearising the second Euler equation leads to
\begin{equation}
\label{euler2}
\begin{split} 
 &\int_{\partial\mathscr{B}_0} \Phi^\mathsf{T}_\mathbf{X} \cdot \mathbf{N} \Phi_\lambda \textrm{d}\Gamma \cdot \delta\widetilde{\boldsymbol\lambda}\\ 
 + &\int_{\partial\mathscr{B}_0} \lambda \left(\mathbf{X}_s-{\boldsymbol\chi}\right) \cdot \left( \textrm{Spin}\left[\frac{\partial\mathbf{X}}{\partial\xi}\right]\cdot\mathbf{B}_\eta - \textrm{Spin}\left[\frac{\partial\mathbf{X}}{\partial\eta}\right]\cdot\mathbf{B}_\xi \right)\, \textrm{d}\xi\textrm{d}\eta \, \delta\widetilde{\mathbf{X}}\\ = &\int_{\partial\mathscr{B}_0} \lambda \Phi^\mathsf{T}_\mathbf{X} \cdot \mathbf{N} \textrm{d}\xi\textrm{d}\eta. 
 \end{split}
\end{equation}
These surface constraint equations are applied for each surface patch independently, i.e. where two surfaces meet, two independent sets of equations with Lagrange multipliers are applied. Moreover, these geometry preserving constraints are not applied to the crack front, since material forces drive the material displacement of those nodes.

\section{Linearised System of Equations and Implementation} \label{system}
A standard linearisation
procedure is applied to the residuals $\mathbf{r}^\textrm{h}_\textrm{m}$,
$\mathbf{r}^\textrm{h}_\textrm{s}$,  $\mathbf{f}^\textrm{h}_\textrm{q}$ \& $r_\lambda$. For the moment, the surface contraints defined in Section \ref{shape_constraints} are excluded.
The global equilibrium solution is obtained using the Newton-Raphson method, solving for the iterative changes in spatial displacements, current material
displacements and the load factor as a fully coupled problem. 
Since the material residual is non-zero only for nodes on the crack front, it is convenient, for the purposes of presentation, to decompose the material nodal positions into those at the crack front $\tilde{\mathbf{X}}_\textrm{f}$, those associated with surfaces $\tilde{\mathbf{X}}_\textrm{s}$ (crack surface and body surfaces) and the rest of the mesh $\tilde{\mathbf{X}}_\textrm{b}$, i.e. $\tilde{\mathbf{X}}=\tilde{\mathbf{X}}_\textrm{f} \cup \tilde{\mathbf{X}}_\textrm{b} \cup \tilde{\mathbf{X}}_\textrm{s}$.
The resulting linear system of equations for iteration $i$ and load step $n+1$ is given as:

\begin{equation} \label{eq::lin_system}
\left[
\begin{array}{ccccc}
\partial_{\widetilde{\mathbf{x}}} \mathbf{r}^\textrm{h}_\textrm{s} &
\partial_\tau \mathbf{r}^\textrm{h}_\textrm{s} &
\partial_{\widetilde{\mathbf{X}}_\textrm{f}} \mathbf{r}^\textrm{h}_\textrm{s} &
\partial_{\widetilde{\mathbf{X}}_\textrm{b}} \mathbf{r}^\textrm{h}_\textrm{s}&
\partial_{\widetilde{\mathbf{X}}_\textrm{s}} \mathbf{r}^\textrm{h}_\textrm{s}\\
0 &
0 &
\partial_{\widetilde{\mathbf{X}}_\textrm{f}} r_\tau &
0 &
0 \\
\partial_{\widetilde{\mathbf{x}}} \mathbf{r}^\textrm{h}_\textrm{m} &
0 &
\partial_{\widetilde{\mathbf{X}}_\textrm{f}} \mathbf{r}^\textrm{h}_\textrm{m} &
0 &
0 \\ 
\partial_{\widetilde{\mathbf{x}}} \mathbf{f}^\textrm{h}_\textrm{q\textunderscore b} &
0 &
\partial_{\widetilde{\mathbf{X}}_\textrm{f}} \mathbf{f}^\textrm{h}_\textrm{q\textunderscore b} &
\partial_{\widetilde{\mathbf{X}}_\textrm{b}} \mathbf{f}^\textrm{h}_\textrm{q\textunderscore b} &
\partial_{\widetilde{\mathbf{X}}_\textrm{s}} \mathbf{f}^\textrm{h}_\textrm{q\textunderscore b} \\  
\partial_{\widetilde{\mathbf{x}}} \mathbf{f}^\textrm{h}_\textrm{q\textunderscore s} &
0 &
\partial_{\widetilde{\mathbf{X}}_\textrm{f}} \mathbf{f}^\textrm{h}_\textrm{q\textunderscore s} &
\partial_{\widetilde{\mathbf{X}}_\textrm{b}} \mathbf{f}^\textrm{h}_\textrm{q\textunderscore s} &
\partial_{\widetilde{\mathbf{X}}_\textrm{s}} \mathbf{f}^\textrm{h}_\textrm{q\textunderscore s} \\  
\end{array}
\right]
\left\{
\begin{array}{c}
\delta\widetilde{\mathbf{x}}^{i+1} \\
\delta\tau^{i+1}\\
\delta\widetilde{\mathbf{X}}_\textrm{f}^{i+1} \\
\delta\widetilde{\mathbf{X}}_\textrm{b}^{i+1} \\
\delta\widetilde{\mathbf{X}}_\textrm{s}^{i+1} 
\end{array}
\right\} = 
-\left[
\begin{array}{l}
\mathbf{r}^\textrm{h}_\textrm{s} \\
r_\tau \\
\mathbf{r}^\textrm{h}_\textrm{m} \\
\mathbf{f}^\textrm{h}_\textrm{q\textunderscore b} \\
\mathbf{f}^\textrm{h}_\textrm{q\textunderscore s}
\end{array}
\right]
\end{equation}
where $\widetilde{\mathbf{X}}^{i+1} = \widetilde{\mathbf{X}}^i+\delta\widetilde{\mathbf{X}}^{i+1}$,  $\widetilde{\mathbf{x}}^{i+1} = \widetilde{\mathbf{x}}^i+\delta\widetilde{\mathbf{x}}^{i+1}$ and $\tau^{i+1} = \tau^i+\delta\tau^{i+1}$. The vectors $\mathbf{f}^\textrm{h}_\textrm{q\textunderscore b}$ and
$\mathbf{f}^\textrm{h}_\textrm{q\textunderscore s}$ are the components of $\mathbf{f}^\textrm{h}_\textrm{q}$ associated with $\widetilde{\mathbf{X}}_\textrm{b}$ and $\widetilde{\mathbf{X}}_\textrm{s}$ respectively. This can be simplified for presentation purposes as:
\begin{equation} \label{eq::lin_system_a}
\left[
\begin{array}{cc}
\mathbf{K}_\textrm{aa} & \mathbf{K}_\textrm{as}\\*[0.5em]
\mathbf{K}_\textrm{sa} & \mathbf{K}_\textrm{ss}
\end{array}
\right]
\left\{
\begin{array}{c}
\delta\widetilde{\mathbf{q}}_\textrm{a}^{i+1} \\*[0.5em]
\delta\widetilde{\mathbf{X}}^{i+1}_\textrm{s} \\*[0.2em]
\end{array}
\right\} = 
-\left[
\begin{array}{l}
\mathbf{f}_\textrm{a}^\textrm{h} \\*[0.5em]
\mathbf{f}_\textrm{q\textunderscore s}^\textrm{h}
\end{array}
\right]
\end{equation}
where $\widetilde{\mathbf{q}}_\textrm{a}$ is the vector of all unknowns excluding the nodal coordinates of the surfaces, $\widetilde{\mathbf{X}}_\textrm{s}$. $\mathbf{f}_\textrm{a}^\textrm{h}$ the corresponding terms on the right hand side of Eq.(\ref{eq::lin_system}). 

In order to preserve the surfaces of the body during the analysis, it is necessary to impose the constraints described in Section \ref{shape_constraints} on the coordinates of the surface nodes, $\widetilde{\mathbf{X}}_\textrm{s}$. Thus, the above system of equations are augmented as follows:
% \begin{equation} \label{eq::lin_system}
% \left[
% \begin{array}{ccc|c}
% \partial_{\widetilde{\mathbf{x}}} \mathbf{r}^\textrm{h}_\textrm{s} &
% \partial_\tau \mathbf{r}^\textrm{h}_\textrm{s} &
% \partial_{\widetilde{\mathbf{X}}_\textrm{f}} \mathbf{r}^\textrm{h}_\textrm{s} &
% \partial_{\widetilde{\mathbf{X}}_\textrm{m}} \mathbf{r}^\textrm{h}_\textrm{s}\\
% 0 &
% 0 &
% \partial_{\widetilde{\mathbf{X}}_\textrm{f}} r_\tau &
% 0 \\
% \partial_{\widetilde{\mathbf{x}}} \mathbf{r}^\textrm{h}_\textrm{m} &
% 0 &
% \partial_{\widetilde{\mathbf{X}}_\textrm{f}} \mathbf{r}^\textrm{h}_\textrm{m} &
% 0 \\ \hline
% \partial_{\widetilde{\mathbf{x}}} \mathbf{f}^\textrm{h}_\textrm{q} &
% 0 &
% \partial_{\widetilde{\mathbf{X}}_\textrm{f}} \mathbf{f}^\textrm{h}_\textrm{q} &
% \partial_{\widetilde{\mathbf{X}}_\textrm{m}} \mathbf{f}^\textrm{h}_\textrm{q} \\  
% \end{array}
% \right]
% \left\{
% \begin{array}{c}
% \delta\widetilde{\mathbf{x}}^{i+1} \\
% \delta\tau^{i+1}\\
% \delta\widetilde{\mathbf{X}}_\textrm{f}^{i+1} \\ \hline
% \delta\widetilde{\mathbf{X}}_\textrm{m}^{i+1} 
% \end{array}
% \right\} = 
% -\left[
% \begin{array}{l}
% \mathbf{r}^\textrm{h}_\textrm{s} \\
% r_\tau \\
% \mathbf{r}^\textrm{h}_\textrm{m} \\ \hline
% \mathbf{f}^\textrm{h}_\textrm{q}
% \end{array}
% \right]
% \end{equation}
\begin{equation} \label{eq::lin_system_b}
\left[
\begin{array}{ccc}
\mathbf{K}_\textrm{aa} & \mathbf{K}_\textrm{as} & 0\\*[0.5em]
\mathbf{K}_\textrm{sa} & \mathbf{K}_\textrm{ss}+\mathbf{B}&\mathbf{C}^\textrm{T}\\*[0.5em]
0&\mathbf{C}+\mathbf{A}&0
\end{array}
\right]
\left\{
\begin{array}{c}
\delta\widetilde{\mathbf{q}}_\textrm{a}^{i+1} \\*[0.5em]
\delta\widetilde{\mathbf{X}}^{i+1}_\textrm{s} \\*[0.5em]
\delta\widetilde{\boldsymbol{\lambda}}^{i+1}
\end{array}
\right\} = 
-\left[
\begin{array}{l}
\mathbf{f}_\textrm{a}^\textrm{h} \\*[0.5em]
\mathbf{f}_\textrm{q\textunderscore s}^\textrm{h}+\mathbf{C}^\textrm{T}\widetilde{\boldsymbol\lambda}\\*[0.5em]
\mathbf{C}(\widetilde{\mathbf{X}}_s-\widetilde{\boldsymbol\chi}_\textrm{s})
\end{array}
\right]
\end{equation}
% \begin{equation}
% \left[ \begin{array}{cc} \partial_{\widetilde{\mathbf{X}}_s}\mathbf{f}^\textrm{h}_{\textrm{q},s} + \mathbf{B} & \mathbf{C}^\mathsf{T} \\ \mathbf{C} + \mathbf{A} & \mathbf{0} \end{array} \right] \left\{ \begin{array}{c} \delta \widetilde{\mathbf{X}} \\ \delta \widetilde{\lambda} \end{array} \right\}= \left[ \begin{array}{c} \mathbf{f}^\textrm{h}_{\textrm{q},2} - \mathbf{C}^\mathsf{T}\widetilde{\lambda} \\ -\mathbf{C}(\widetilde{\mathbf{X}}_s-\widetilde{\boldsymbol\chi}_\textrm{s}) \end{array} \right]
% \end{equation}
where
\begin{equation}
\mathbf{C}= 
\A_\textrm{TRI} 
\int_{\textrm{TRI}}
{\boldsymbol\Phi}_\lambda^\mathsf{T} \mathbf{N} \cdot {\boldsymbol\Phi}_\mathbf{X} \textrm{d}\xi\textrm{d}\eta,
\end{equation}
\begin{equation}
\mathbf{B}= 
\A_\textrm{TRI} 
\int_{\textrm{TRI}}
\lambda {\boldsymbol\Phi}^\mathsf{T}_\mathbf{X} \left( \textrm{Spin}\left[\frac{\partial\mathbf{X}}{\partial\xi}\right]\cdot\mathbf{B}_\eta - \textrm{Spin}\left[\frac{\partial\mathbf{X}}{\partial\eta}\right]\cdot\mathbf{B}_\xi \right) \textrm{d}\xi\textrm{d}\eta
\end{equation}
and
\begin{equation}
\mathbf{A}= 
\A_\textrm{TRI} 
\int_{\textrm{TRI}}
{\boldsymbol\Phi}^\mathsf{T}_\lambda \left(\mathbf{X}_\textrm{s}-{\boldsymbol\chi}_\textrm{s}\right) \cdot \left( \textrm{Spin}\left[\frac{\partial\mathbf{X}}{\partial\xi}\right]\cdot\mathbf{B}_\eta - \textrm{Spin}\left[\frac{\partial\mathbf{X}}{\partial\eta}\right]\cdot\mathbf{B}_\xi \right)
\textrm{d}\xi\textrm{d}\eta.
\end{equation}

%It is important to state that the crack extension criterion is evaluated, and the crack direction is determined, using the configurational forces derived from the system of equations before $\mathbf{f}^\textrm{h}_\textrm{q}$ is added to the system of equations, thereby only having an impact on the stability of the solution and not on the crack propagation criterion. 
When solvng this nonlinear system of equations, convergence is quadratic and typically requires 3-4 iterations per load step to achieve convergence.
We adopt an Total Arbitrary Lagrangian Eulerian approach, and at beginning of each new load step the current material mesh becomes the new reference mesh at the next iteration.
Discretisation is undertaken using 3D tetrahedral elements with hierarchical basis functions of arbitrary polynomial order~\cite{R12} in the spatial domain. A linear approximation is adopted in the material domain.

%An adaptive mesh refinement strategy is adopted that includes both h-refinement, using an edge-based splitting algorithm \cite{R14}, and p-refinement using hierarchic basis functions identified above~\cite{R12}. First, all elements $\mathcal{E}_1$ adjacent to the crack
%front are selected, then a larger set of elements $\mathcal{E}_2$
%adjacent to and including set $\mathcal{E}_1$ are selected. All elements in set
%$\mathcal{E}_2$ are subject to edge-based splitting. The spatial nodal positions are discretised using higher-order approximations in the vicinity of the crack front. The spatial nodal positions in all elements in 
%$\mathcal{E}_2 \backslash \mathcal{E}_1$ are discretised using second-order approximation functions and elements in 
%$\mathcal{E}_1$ using third-order. This refinement strategy is automated so that the mesh is locally refined while the crack is propagating. This is illustrated in Fig~\ref{F4a}. As the crack front moves forward, elements are removed from set $\mathcal{E}_2$ and elements revert to their original state.

The solution strategy presented in this paper is implemented for parallel shared memory computers, utilising open-source libraries. MOAB,  
a mesh-oriented database \cite{R18}, is used to store mesh data, including input and output operations and information about mesh topology. 
PETSc (Portable, Extensible Toolkit for Scientific Computation \cite{R19}) is used for parallel
matrix and vector operations, the solution of linear system of equations and other algebraic operations. MOAB and PETSc are integrated in MoFEM \cite{R_mofem} where approximation base, finite elements are implemented.

\section{Numerical Examples}

Three numerical examples of crack propagation in brittle materials are presented. The first two examples presented consider the fracture of unirradiated nuclear graphite. The third example considers the fracture of PMMA.

\subsection{Graphite cylinder slice test}
\begin{figure}
\begin{center}
\includegraphics[width=120mm]{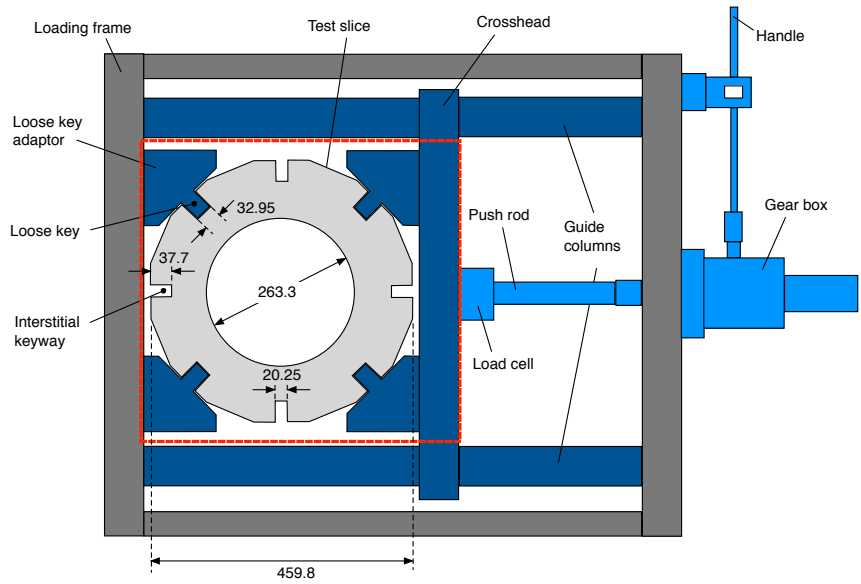}
\end{center}
\caption{Experimental loading frame for graphite cylinder slice test. All dimensions in mm. \label{brick_loading}}
\end{figure}

This numerical example considers a slice of a graphite cylindrical brick, placed in a loading rig and loaded as shown in Figure~\ref{brick_loading}. The red box indicates the part modelled in the numerical analysis. The loose key adaptors on the left are fully fixed along their left hand side. The numerical load is applied to the mid-point of the crosshead beam. The brick slice is 25 mm thick, the Young's modulus is 10,900 MPa, Poisson's ratio is 0.2 and Griffith energy is 0.23 N/mm. The specimen is loaded via the key adaptors. Contact between the key adaptors and the graphite slice only occurs on one edge of the loose keyway and this is modelled by tied degrees of freedom on those edges.

\begin{figure}
\begin{center}
\includegraphics[width=120mm]{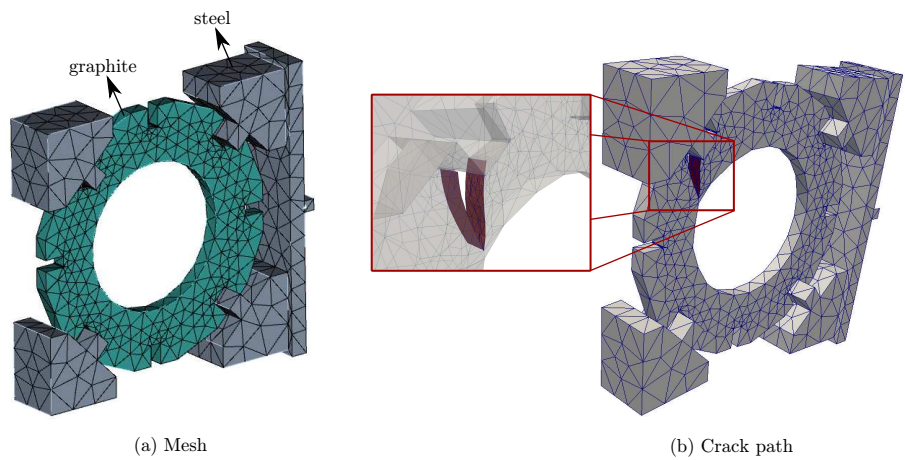}
\end{center}
\caption{Geometry and crack path for graphite cylinder slice test. \label{slice_mesh_and_CrackPath}}
\end{figure}

\begin{figure}
\begin{center}
\includegraphics[width=120mm]{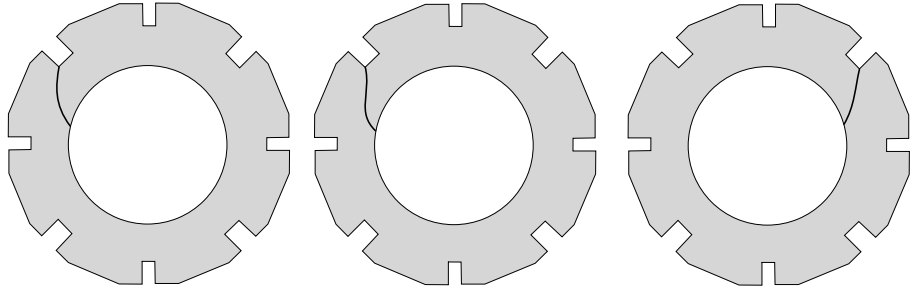}
\end{center}
\caption{Experimentally determined crack paths for graphite cylinder slice test. \label{slice_crack_paths}}
\end{figure}

\begin{figure}
\begin{center}
\includegraphics[width=120mm]{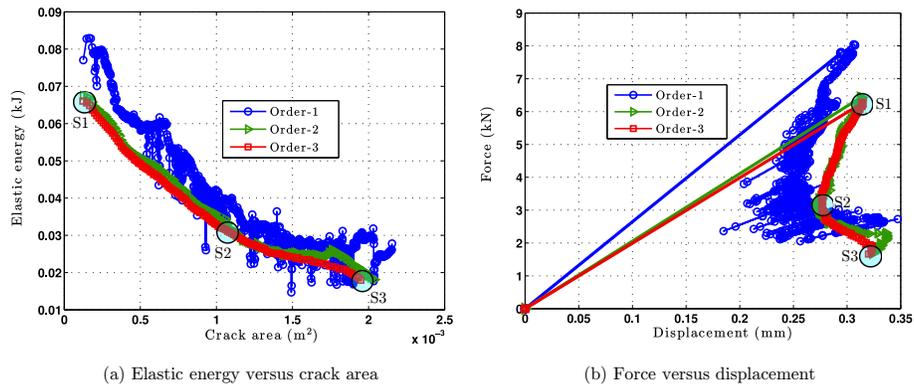}
\end{center}
\caption{Energy versus crack area and load versus displacement plots for graphite cylinder slice test. \label{slice_EA_and_LD}}
\end{figure}

The mesh for this example is shown in Figure~\ref{slice_mesh_and_CrackPath}(a) and is consists of 6033 tetrahedral elements. The numerical analyses were undertaken using one mesh but repeated for 1st-order, 2nd-order and 3rd-order approximations. The numerically predicted crack path is shown in Figure~\ref{slice_mesh_and_CrackPath}(b). It can be seen that the crack propagates from the keyway corner with a curved trajectory to the free surface of the inner bore. These compare well with the experimentally determined crack paths shown in Figure~\ref{slice_crack_paths}.
%, against which the numerically predicted stiffness and ultimate load also compare very well. 
The elastic energy versus the crack area and the force versus displacement plots are shown in Figures \ref{slice_EA_and_LD}(a) and \ref{slice_EA_and_LD}(b) respectively. The displacement in Figure \ref{slice_EA_and_LD}(b) is known as the generalised displacement and does not represent a particular point on the structure, but its value is work conjugate to the applied forces and is calculated as $u=2\Psi / \lambda f$, where $f=1N$ is the reference force, $\Psi$ is the elastic energy, $\lambda$ is the load factor and $u$ is the generalised displacement. The arc-length control, described earlier, is used to trace the nonlinear response. 

% DO WE INCLUDE THESE NEXT TWO SENTENCEs?
% The experimental ultimate load was 6.2 kN. The numerical ultimate loads were 7.79 kN, 6.42 kN, 6.25 kN for the 1st-order, 2nd-order, 3rd-order analyses respectively as shown in Figure \ref{slice_EA_and_LD}(b). 

\begin{figure}
\begin{center}
\includegraphics[width=120mm]{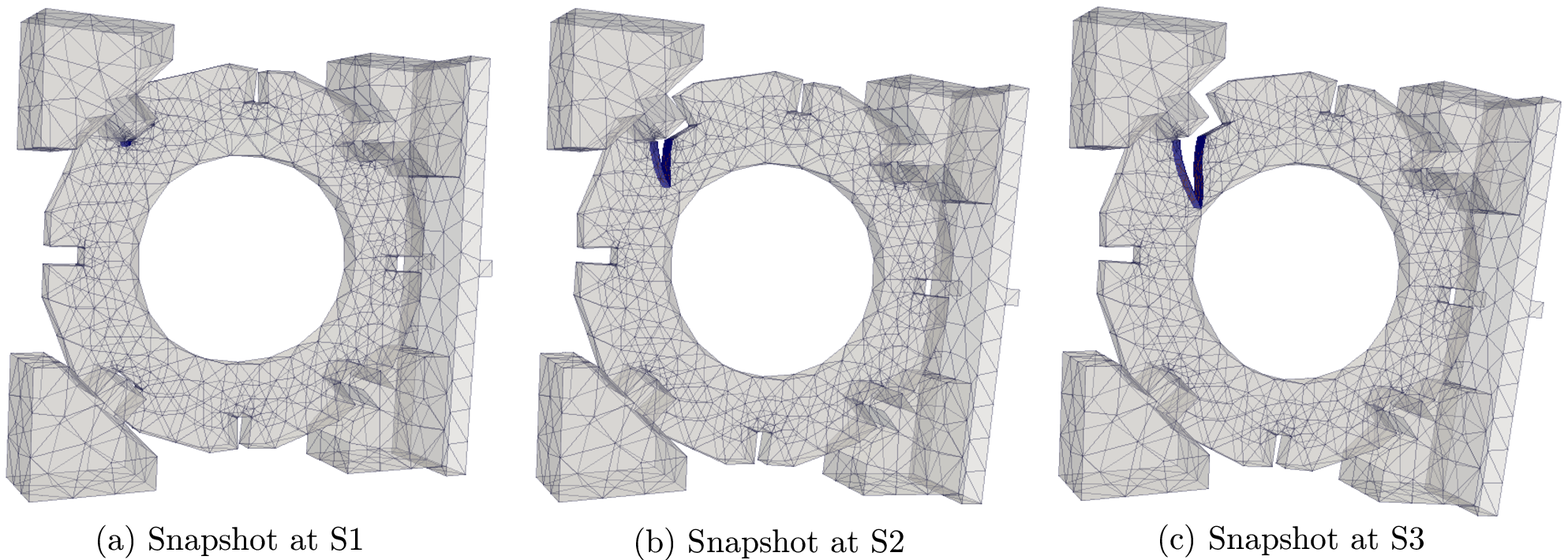}
\end{center}
\caption{Snap shots for points S1, S2 and S3 shown in Figure~\ref{slice_EA_and_LD} for graphite cylinder slice test. \label{slice_snap_shots}}
\end{figure}

%The crack surface is a result of maximum release energy, therefor any other shape of crack surface area would release less energy. The energy release rate is assumed constant and is a material parameter ($g_c$). 
Snapshots of brick slice at three points S1, S2 and S3, shown in Figure \ref{slice_EA_and_LD} are shown in Figure \ref{slice_snap_shots}(a), \ref{slice_snap_shots}(b) and \ref{slice_snap_shots}(c) respectively. It should be noted that Figure \ref{slice_EA_and_LD} demonstrates that, as compared to the fluctuating results for 1st-order analysis, the results for 2nd-order and 3rd-order analyses are very smooth. The fluctuation in the results for the 1st order analysis is due to the shear locking. 

\subsection{Mixed mode I--III loading in three-point bending test}
\begin{figure}
\begin{center}
\includegraphics[width=120mm]{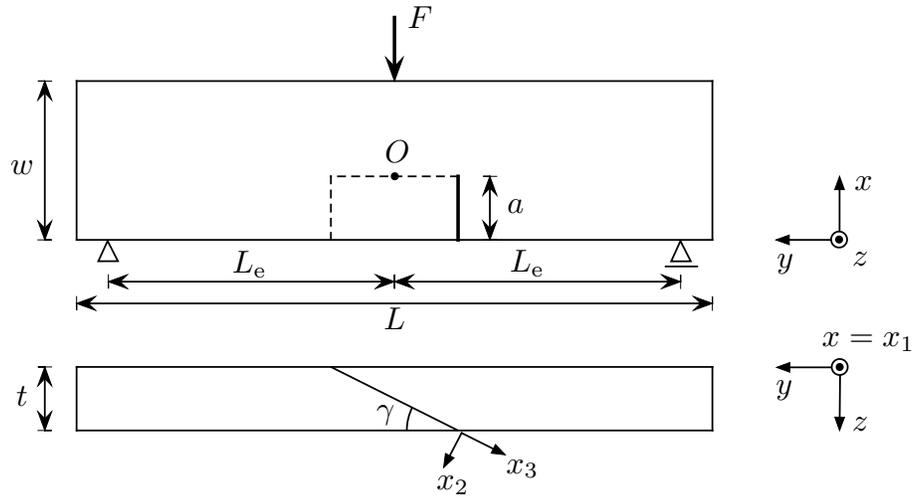}
\end{center}
\caption{Mixed mode I--III crack growth in PMMA
specimen loaded in a three-point bending configuration~\cite{R29}. The specimen length is $L=260$mm, $L_\textrm{e}=120$mm, depth $w=60$mm, thickness $t=10$mm, notch height $a=20$mm and notch inclination $\gamma=45^\circ$, $60^\circ$ and $75^\circ$. \label{lazarus_problem}}
\end{figure}

\begin{figure}
\begin{center}
\includegraphics[width=100mm]{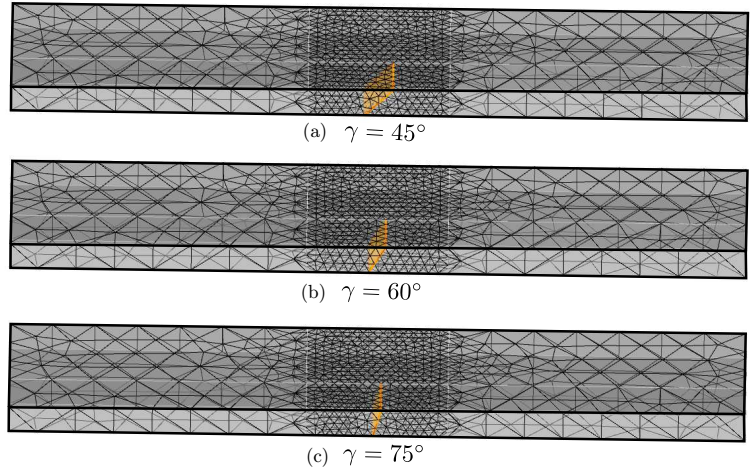}
\end{center}
\caption{Meshes for mixed mode loading example.\label{lazarus_meshes}}
\end{figure}

The second example assesses the ability of the proposed approach to simulate the mixed mode I--III experiments of \cite{R29} on PMMA pre-notched beams loaded in a three-point-bending configuration. This problem was also simulated by Pandolfi and Ortiz~\cite{R30} (although they did not model the entire specimen, instead restricting calculations to the central section of the specimen, applying  equivalent static boundary conditions representative of three-point bending). The specimens are $260$mm long, $60$mm deep, and $10$mm thick, pre-cracked at $45^\circ$, $60^\circ$, and $75^\circ$ to the front face, Figure~\ref{lazarus_problem}. Consistent with~\cite{R29}, Young's modulus $E=2800$MPa, Poisson's ratio $\nu=$0.38, and Griffith energy 0.352 N/mm are used. The inclination of the notch results in an initial mixed mode I--III loading of the crack front. The experiments demonstrated that for homogeneous and linear-elastic isotropic materials the growing crack, loaded in mixed I--III mode, reorientates toward a pure mode I situation. The meshes are shown in Figure \ref{lazarus_meshes} and consist of 8696, 7824 and 8705 tetrahedral elements for $45^\circ$, $60^\circ$, and $75^\circ$ initial notch angles respectively. 

\begin{figure}
\begin{center}
\includegraphics[width=120mm]{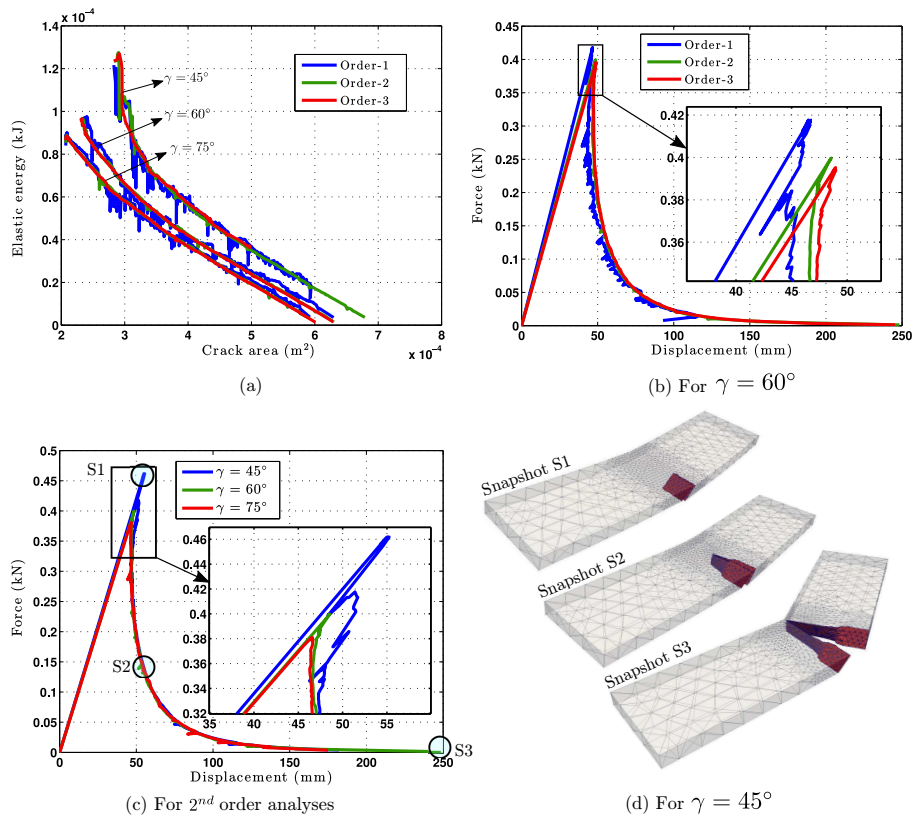}
\end{center}
\caption{Elastic energy versus crack area and force versus displacement plots for 1st-, 2nd- and 3rd-order of approximation and various initial crack inclinations for mixed mode loading example. \label{lazarus_plots}}
\end{figure}

\begin{figure}
\begin{center}
\includegraphics[width=80mm]{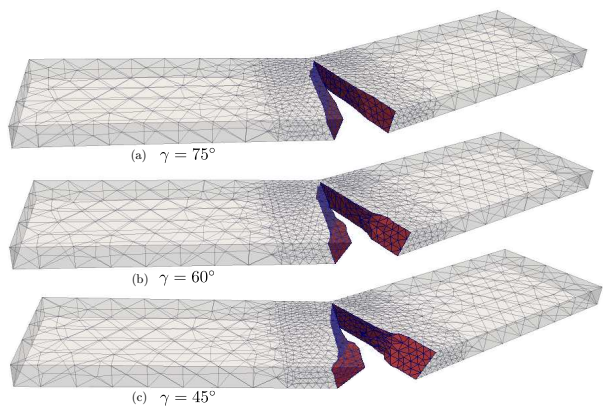}
\end{center}
\caption{Final crack surfaces for various pre-crack inclinations for mixed mode loading example. \label{lazarus_cracks}}
\end{figure}

All three cases are solved for 1st, 2nd and 3rd-order of approximation. Elastic energy versus crack area plots are shown in Figure \ref{lazarus_plots}(a). Consistent with the previous example, the response for all three cases fluctuates for the 1st-order analysis but are smooth for the 2nd and 3rd-order analyses. 
Figure \ref{lazarus_plots}(b) compares the load-displacement response for $\gamma=60^\circ$, for different orders of approximation. Figure \ref{lazarus_plots}(c) compares the load-displacement response for 2nd-order approximations, for all three initial notch angles. The ultimate load for $\gamma=45^\circ$, $\gamma=60^\circ$ and $\gamma=75^\circ$ are 0.46 kN, 0.4 kN and 0.38 kN respectively, decreasing with increasing initial notch angle. 
For the $\gamma=45^\circ$ case, snapshots of the evolving crack for the 2nd-order analyses associated with points S1, S2 and S3 (see Figure \ref{lazarus_plots}(c)) are shown in Figure \ref{lazarus_plots}(d). Furthermore, the final crack surfaces for $\gamma=75^\circ$, $\gamma=60^\circ$ and $\gamma=45^\circ$ are shown in Figure \ref{lazarus_cracks}(a), \ref{lazarus_cracks}(b) and \ref{lazarus_cracks}(c) respectively, which clearly shows the re-orientation towards a mode I situation. Table \ref{lazarus_table} compares the experimentally measured average kink angles, obtained by averaging the kink angles on the front and back surfaces of the specimen, and the corresponding numerical values. The numerical results capture the general trend of increasing kink angle with increasing pre-crack inclination and mode III component. The difference between the numerical and experimental results are very small.

\begin{table}
\begin{center}
\begin{tabular}{c|c|c|c} \hline
$\gamma$ & Experiment \cite{R29} & Numerical & Error \%  \\ \hline
$45^\circ$ & $61.9^\circ$ & $56.02^\circ$ & $9.49$ \\ 
$60^\circ$ & $38.4^\circ$ & $37.46^\circ$ & $2.45$ \\
$75^\circ$ & $21.1^\circ$ & $20.87^\circ$ & $1.09$ \\ \hline
\end{tabular}
\end{center}
\caption{Comparison of experimental~\cite{R29} and numerical initial kink angle for different notch inclinations. \label{lazarus_table}}
\end{table}

\subsection{Torsion test}
\begin{figure}
\begin{center}
\includegraphics[width=140mm]{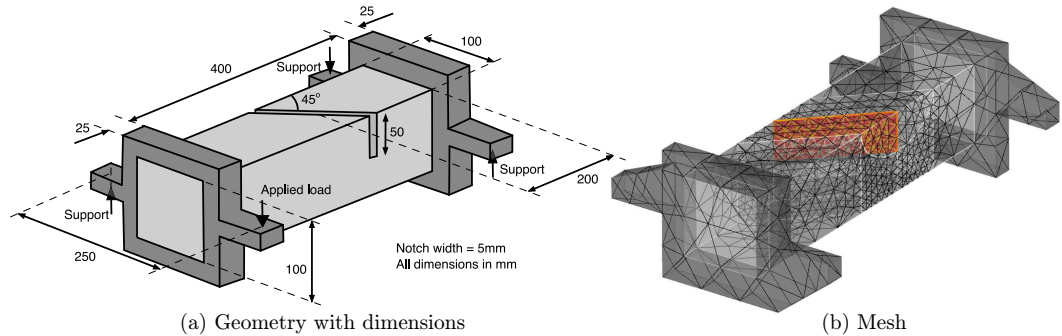}
\end{center}
\caption{Geometry and mesh for torsion test. \label{torsion_geo_mesh}}
\end{figure}

An experimental study by Brokenshire~\cite{R23} of a torsion test of a plain concrete notched prismatic beam ($400\textrm{mm}\times 100\textrm{mm}\times 100\textrm{mm}$) has been repeated for nuclear graphite. The experimental procedure and full details of the boundary conditions and dimensions for the original study are described in Jefferson et al.~\cite{R22} and illustrated in Figure~\ref{torsion_geo_mesh}(a). The notch is placed at an oblique angle across the beam and extends to half the depth. The beam is placed in a steel loading frame, supported at three corners and loaded at the fourth corner. The material properties used for this example is the same as used in the graphite cylinder slice test example. 

\begin{figure}
\begin{center}
\includegraphics[width=120mm]{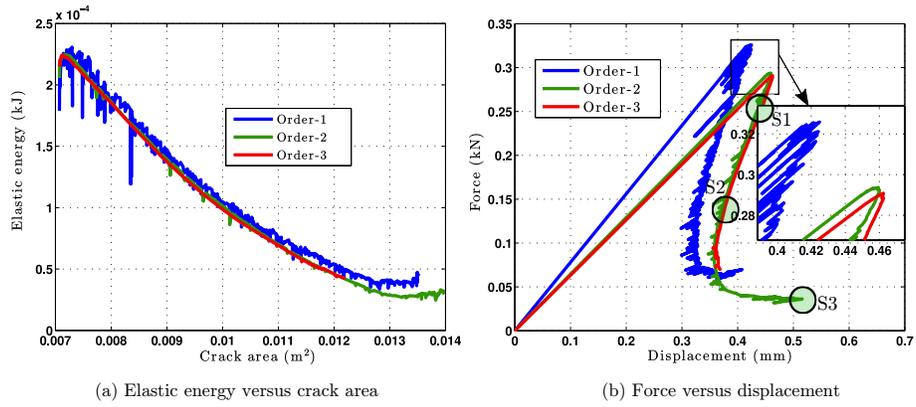}
\end{center}
\caption{(a) Elastic energy versus crack area and (b) load versus displacement plots for torsion test . \label{torsion_EvsA_and_FvsU}}
\end{figure}

\begin{figure}
\begin{center}
\includegraphics[width=120mm]{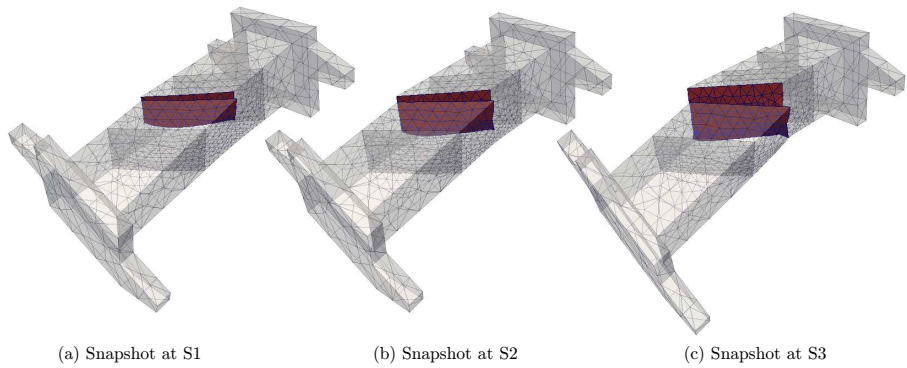}
\end{center}
\caption{Snap shots for points S1, S2 and S3 shown in Figure~\ref{torsion_EvsA_and_FvsU} for torsion test. \label{torsion_snapshots}}
\end{figure}

\begin{figure}
\begin{center}
\includegraphics[width=120mm]{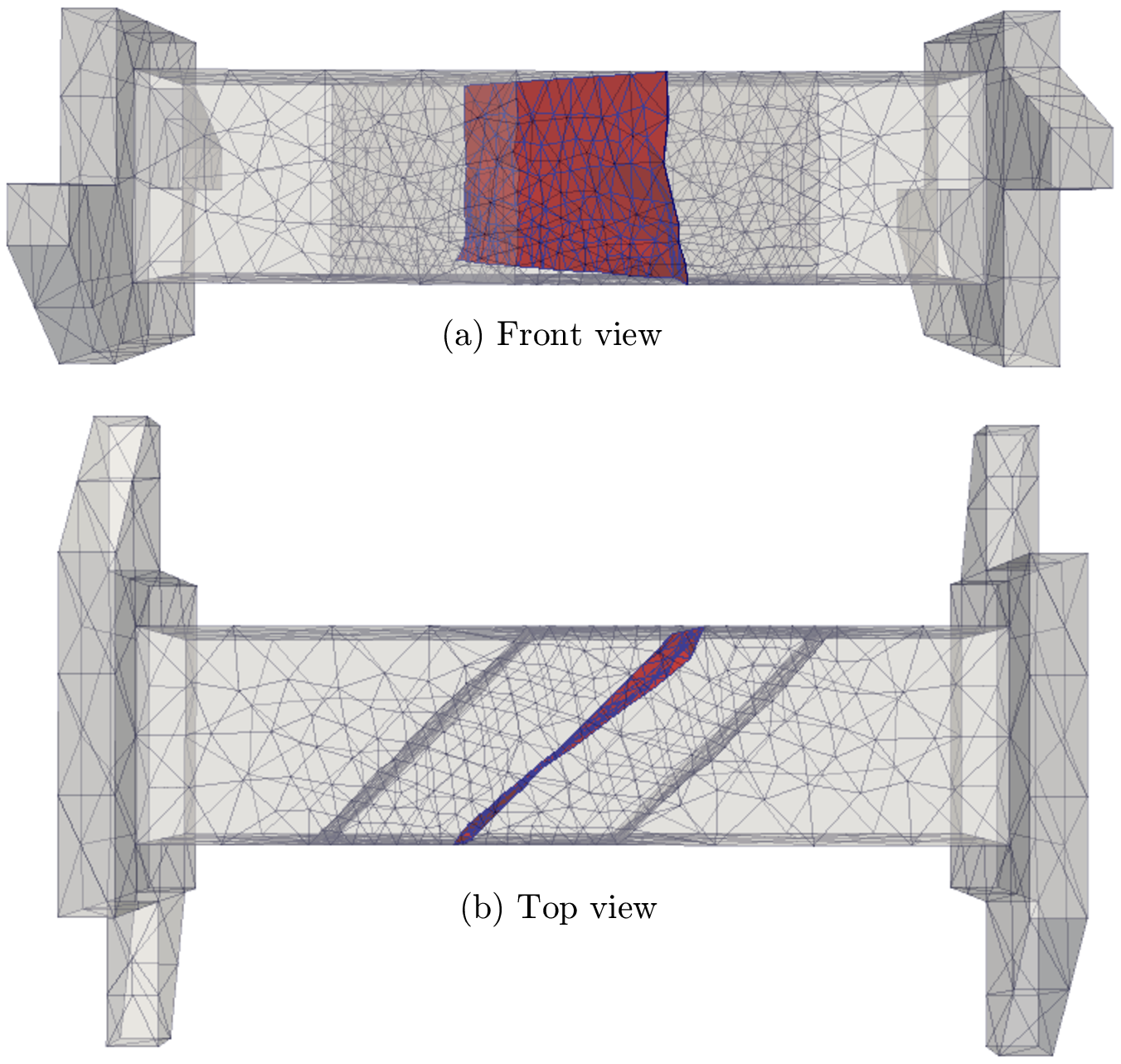}
\end{center}
\caption{The simulated crack surface viewed from front and top for torsion test. \label{torsion_crack_surface_view}}
\end{figure}

The beam and steel frame are discretised using tetrahedral elements and is shown in Figure~\ref{torsion_geo_mesh}(b). The mesh consists of 11890 elements and the problem is solved with 1st-order, 2nd-order and 3rd order of approximation. Figure \ref{torsion_EvsA_and_FvsU}(a) and  \ref{torsion_EvsA_and_FvsU}(b) shows the elastic energy-crack area and load-displacement response for the three numerical tests respectively. Good numerical convergence is observed with increasing order of approximation and the arc-length control method is able to track the dissipative loading path. The ultimate load for 1st-order, 2nd-order and 3rd-order of approximation is 0.326 kN, 0.294 kN and 0.291 kN respectively.  The snapshots of evolving crack at points S1, S2 and S3 (see Figure \ref{torsion_EvsA_and_FvsU}(b)), are shown in Figure \ref{torsion_snapshots}(a),  \ref{torsion_snapshots}(b) and  \ref{torsion_snapshots}(c) respectively. The front and top view of the simulated crack surface is also shown in Figure \ref{torsion_crack_surface_view}(a) and \ref{torsion_crack_surface_view}(b) respectively, which clearly shows the complicated crack path. It is worth noting that the same problem was discussed in our previous paper~\cite{R31} but the material was concrete rather than graphite. In this previous case the crack path showed excellent agreement with experiments but over predicted the experimental ultimate load by approximately $2.5$ times. This was a consequence of assuming linear elastic fracture mechanics for a problem where the size of the fracture process zone is significant compared to the size of the problem. This is not an issue in the current situation because graphite's microstructure is significantly smaller than for concrete and the assumption of linear elastic fracture mechanics is valid.

\section{Conclusions}

A formulation for brittle fracture in 3D elastic solids within the context of configurational mechanics has been presented. Configurational forces are the driving force for advancement of the crack front. The local form of the first law of thermodynamics provides a condition for equilibrium of the crack front and the direction of crack propagation is given by the direction of the configurational forces on the crack front. This crack advancement maximises the local dissipation. The moving crack front is continuously resolved by the finite element mesh, without the need for face splitting or the use of enrichment techniques. 

A monolithic solution strategy has been described that simultaneously solves for both the material displacements (i.e. crack extension) and the spatial displacements. In order to trace the dissipative loading path, an arc-length procedure has been developed that controls the incremental crack area growth. In order to maintain mesh quality, smoothing of the mesh is undertaken as a continuous process, together with face flipping, node merging and edge splitting where necessary. Hierarchical basis functions of arbitrary polynomial order are adopted to increase the order of approximation without the need to change the finite element mesh. 

Three numerical examples have been presented to demonstrate both the accuracy and robustnesss of the formulation. Convergence studies have been undertaken in all cases. All three problems demonstrate the ability to simulate experimental crack paths. 

\section*{Acknowledgements} This work was supported by EDF Energy Nuclear
Generation Ltd and The Royal Academy of Engineers. The views expressed in this paper are those of the authors and not necessarily those of EDF Energy Nuclear Generation Ltd. Analyses were undertaken using the EPSRC funded ARCHIE-WeSt High Performance 
Computer (www.archie-west.ac.uk). EPSRC grant no. EP/K000586/1.

\section*{References}

%\bibliography{mybibfile}

\end{document}